\documentclass[journal=jctcce,manuscript=article]{achemso}

\usepackage{amssymb,amsmath,multicol,multirow,longtable,array,mathpazo}
\usepackage{mhchem}
\usepackage{makecell}
\usepackage{enumerate}
\usepackage[T1]{fontenc} 

\usepackage{threeparttable}
\usepackage{color}
\usepackage{graphicx}
\usepackage{longtable}
\usepackage{enumerate}
\usepackage{mathrsfs}
\usepackage{amsfonts}
\usepackage{amsmath}
\usepackage{amssymb}
\usepackage{stmaryrd}
\usepackage{dcolumn}
\usepackage{bm}
\usepackage{bbm}
\usepackage{subfigure}
\usepackage{pgfplots}
\usepackage{verbatim}
\usepackage{accents}
\usepackage{diagbox}
\usepackage{simplewick} 

\title{$\mathbbm{i}$CAS: Imposed Automatic Selection and Localization of Complete Active Spaces}

\author{Yibo Lei}
\affiliation{Key Laboratory of Synthetic and Natural Functional Molecule of the Ministry of Education, College of Chemistry \& Materials Science, Shaanxi key Laboratory of Physico-Inorganic Chemistry, Northwest University, Xi'an 710127, Shaanxi, P. R. China}

\author{Bingbing Suo}
\affiliation{Institute of Modern Physics, Northwest University, and Shaanxi Key Laboratory for Theoretical Physics Frontiers, Xi'an 710127, Shaanxi, P. R. China}

\author{Wenjian Liu}
\affiliation{Qingdao Institute for Theoretical and Computational Sciences, Institute of Frontier and Interdisciplinary Science, Shandong University, Qingdao, Shandong  266237, P. R. China}
\email{liuwj@sdu.edu.cn}

\begin{document}

\begin{abstract}
It is shown that, in the spirit of ``from fragments to molecule'' for localizing molecular orbitals [J. Chem. Theory Comput. 7, 3643 (2011)],
a prechosen set of occupied/virtual valence/core atomic/fragmental orbitals can be transformed to an equivalent
 set of localized occupied/virtual pre-molecular orbitals (pre-LMO), which
can then be taken as probes to select the same number of maximally matching
localized occupied/virtual Hartree-Fock (HF) or restricted open-shell Hartree-Fock (ROHF)
 molecular orbitals as the initial local orbitals spanning the desired complete active space (CAS). In each cycle of the self-consistent field (SCF) calculation,
the CASSCF orbitals can be localized by means of the noniterative ``top-down least-change'' algorithm for localizing ROHF
orbitals [J. Chem. Phys. 2017, 146, 104104], such that the maximum matching between the orbitals of two adjacent iterations can readily be monitored,
leading finally to converged localized CASSCF orbitals that overlap most the guess orbitals.
Such an approach is to be dubbed as ``imposed CASSCF'' ($\mathbbm{i}$CASSCF or simply $\mathbbm{i}$CAS in short)
for good reasons: (1) it has been assumed that only those electronic states that have largest projections onto
the active space defined by the prechosen atomic/fragmental orbitals are to be targeted. This is certainly an imposed constraint
but has wide applications in organic and transition metal chemistry where valence (or core) atomic/fragmental orbitals can readily be identified.
(2) The selection of both initial and optimized local active orbitals is imposed from the very beginning
by the pre-LMOs (which span the same space as the prechosen atomic/fragmental orbitals). Apart from the (imposed) automation and localization,
$\mathbbm{i}$CAS has an additional merit: the CAS is guaranteed to be the same for all geometries for
the pre-LMOs do not change in character with geometry.
Both organic molecules and transition metal complexes are taken as showcases to reveal the efficacy of $\mathbbm{i}$CAS.
\end{abstract}

\maketitle

\clearpage
\newpage

\section{Introduction}\label{sec:intro}
The common paradigm for quantum chemical descriptions of
 strongly correlated systems of electrons is to decompose the overall correlation into static and dynamic components. The former
originates from a set of strongly coupled configurations whereas the latter stems from the remaining, weakly coupled ones.
Transition metal complexes with multiple unpaired electrons and bond formation/cleavage are typical examples of strongly correlated systems.
Yet, it is hardly possible to select manually the full set of strongly coupled many-electron configurations,
apart from a few of them that can be guessed by virtue of chemical or physical intuitions. One way out is to preselect a small number ($n_o$) of frontier Hartree-Fock (HF) or
restricted open-shell Hartree-Fock (ROHF) orbitals and then allow
$n_e$ electrons to occupy such \emph{active} orbitals in all possible ways, thereby leading to a complete active space (CAS) that is usually denoted as CAS$(n_e, n_o$).
If the space is directly diagonalized to account for interactions between the configurations, one would have a simple CASCI approach, with CI standing for configuration interaction.
However, it is often the case that HF/ROHF orbitals are very poor for correlation, such that it is necessary to reoptimize them along with the CI coefficients or simply replace them
with other type of orbitals (e.g., natural orbitals (NO) generated in one way or another\cite{NOON-UHFa,NOON-UHFb,NOON-UHFc,NOON-CISD}).
The former leads to the so-called CASSCF method\cite{FORS1978,CASSCF,WernerRev1987,CASSCFRev1987,MCSCFRev1987,MCSCFRev1998,MCSCFrev2012}, which has
long been the cornerstone of\emph{ ab initio} quantum chemistry for strong correlation thanks to its operational simplicity
as well as ever improved algorithms\cite{Co-Iter,NewCAS2019,NewCAS2020}. Nevertheless, CASSCF
has been plagued by several factors: (1) it is by no means trivial to select which and how many orbitals as active orbitals. (2) It is often the case
that some desired active orbitals run out of the CAS during the self-consistent field (SCF) iterations, thereby resulting in undesired solutions or
no convergence. (3) It is
particularly difficult to maintain the same CAS when scanning potential energy surfaces. (4) The size of CAS grows combinatorially with respect to $n_o$ and $n_e$,
to name just the major ones. While the combinatorial complexity can be alleviated to a large extent by imposing certain restrictions on the occupation patterns so as to \emph{a priori} reduce
the space size\cite{GVB,CCASSCF,GVBCAS1996,QCASSCF,rCIa,rCIb,ORMAS,RASSCFa,RASSCFb,GAS2011,splitGAS,splitGAS2015,ASD-CASSCF} or by using some highly efficient \emph{posteriori} selection schemes as the CI solver\cite{v2RDM2008,v2RDM2016,DMRGSCF2008a,DMRGSCF2008b,DMRGSCF2009,DMRGSCF2013,DMRGSCF2014,DMRGSCF2017a,DMRGSCF2017b,FCIQMCSCF2015,FCIQMCSCF2016,HBCISCF2017,HBCISCF2021,iCASSCF2019,ASCISCF,ASCISCF2,iCISCF},
the first three issues pertinent to orbital selection and optimization are more delicate. Although the selection of active orbitals can be automated by
using, e.g, occupation numbers of NOs\cite{NOON-UHFa,NOON-UHFb,NOON-UHFc,NOON-CISD,NOON-MP2,NOON-NEVPT2}, information entropies\cite{OrbitalEntropya,OrbitalEntropyb,autoCAS},
machine learning\cite{ML-auto}, subspace projection\cite{AVAS,PiOS,GVB2018}, stepwise testing\cite{ABC}, etc., the so-constructed CAS
cannot be guaranteed to be the same for all geometries of complex systems due to the underlying cutoff thresholds or varying parameters.
Herewith, we propose an automated approach for constructing a CAS that is guaranteed to be the same for all geometries. Like the subspace projection approaches\cite{AVAS,PiOS},
it is based on the (minimal) assumption that the electronic states of interest have largest projections onto a prechosen subspace. However, our approach originates actually from the generic idea of ``from fragments to molecule'' (F2M) for constructing localized molecular orbitals (LMO)\cite{F2M,ACR-FLMO,Triad,OpenFLMO},
by regarding atoms also as fragments. A set of pre-LMOs can readily be constructed from the prechosen atomic/fragmental orbitals
and are then used as probes to
select automatically initial local active orbitals. The CASSCF orbitals can also be localized in each iteration so as to facilitate the matching of
orbitals between two adjacent iterations. Since  the selection of both initial and optimized local active orbitals is imposed from the very beginning
by the pre-LMOs (which span the same space as the prechosen atomic/fragmental orbitals), this approach will be dubbed as ``imposed CASSCF'' ($\mathbbm{i}$CASSCF or simply $\mathbbm{i}$CAS in short).
It should not be confused with the iCASSCF approach\cite{iCASSCF2019} that solves the CAS problem in terms of increments.
The algorithms are described in detail in Sec. \ref{iCAS}. The efficacy of $\mathbbm{i}$CAS
will be revealed in Sec. \ref{Results} by taking both organic molecules and transition metal complexes as examples.
The presentation is closed with concluding remarks Sec. \ref{Conclusion}.

\section{$\mathbbm{i}$CAS}\label{iCAS}
The proposed $\mathbbm{i}$CAS approach involves two essential ingredients, i.e., preparation of pre-LMOs
as probes to select initial local active orbitals and subspace matching in each iteration.
The former determines the size and hence quality of the CAS by virtue of chemical/physical intuitions, whereas the latter
dictates the converge by matching the orbitals of two adjacent iterations.
\subsection{pre-LMO}\label{preLMO}
To introduce the concept of pre-LMO, it is necessary to briefly recapitulate the F2M scheme\cite{F2M,ACR-FLMO,Triad,OpenFLMO} for constructing ROHF LMOs.
It involves two steps: (1) generate a set of orthonormal primitive fragment localized molecule orbitals (pFLMO) $\{\phi_{\mathrm{c}}^{\mathrm{pFLMO}},\phi_{\mathrm{a}}^{\mathrm{pFLMO}},\phi_{\mathrm{v}}^{\mathrm{pFLMO}}\}$
from subsystem SCF calculations and localizations (for more details, see Ref.\cite{F2M,ACR-FLMO});
(2) perform a least-change type of block-diagonalization of the converged Fock matrix represented in the pFLMO basis. Since the pFLMOs
are already classified into doubly occupied (denoted by c), singly occupied (denoted by a) and unoccupied (denoted by v), the ROHF equation takes the following form
\begin{equation}
\begin{pmatrix}
\mathbf{F}_{\mathrm{cc}} & \mathbf{F}_{\mathrm{ca}}&\mathbf{F}_{\mathrm{cv}} \\
\mathbf{F}_{\mathrm{ac}} & \mathbf{F}_{\mathrm{aa}}&\mathbf{F}_{\mathrm{av}} \\
\mathbf{F}_{\mathrm{vc}} & \mathbf{F}_{\mathrm{va}}&\mathbf{F}_{\mathrm{vv}}
\end{pmatrix}
\begin{pmatrix}
\mathbf{C}_{\mathrm{cc}} & \mathbf{C}_{\mathrm{ca}}&\mathbf{C}_{\mathrm{cv}} \\
\mathbf{C}_{\mathrm{ac}} & \mathbf{C}_{\mathrm{aa}}&\mathbf{C}_{\mathrm{av}}\\
\mathbf{C}_{\mathrm{vc}} & \mathbf{C}_{\mathrm{va}}&\mathbf{C}_{\mathrm{vv}}
\end{pmatrix}=
\begin{pmatrix}
\mathbf{C}_{\mathrm{cc}} & \mathbf{C}_{\mathrm{ca}}&\mathbf{C}_{\mathrm{cv}} \\
\mathbf{C}_{\mathrm{ac}} & \mathbf{C}_{\mathrm{aa}}&\mathbf{C}_{\mathrm{av}}\\
\mathbf{C}_{\mathrm{vc}} & \mathbf{C}_{\mathrm{va}}&\mathbf{C}_{\mathrm{vv}}
\end{pmatrix}
\begin{pmatrix}
\mathbf{E}_{\mathrm{c}} & \mathbf{0} &\mathbf{0}     \\
\mathbf{0}      & \mathbf{E}_{\mathrm{a}}&\mathbf{0}\\
\mathbf{0}      & \mathbf{0} &\mathbf{E}_{\mathrm{v}}
\end{pmatrix}. \label{F:CMO3}
\end{equation}
After diagonalization to obtain the canonical molecular orbitals (CMO)
\begin{eqnarray}
\{\psi_{\mathrm{c}}^{\mathrm{CMO}},\psi_{\mathrm{a}}^{\mathrm{CMO}},\psi_{\mathrm{v}}^{\mathrm{CMO}}\}=(\phi_{\mathrm{c}}^{\mathrm{pFLMO}}, \phi_{\mathrm{a}}^{\mathrm{pFLMO}}, \phi_{\mathrm{v}}^{\mathrm{pFLMO}})
\begin{pmatrix}
\mathbf{C}_{\mathrm{cc}} & \mathbf{C}_{\mathrm{ca}}&\mathbf{C}_{\mathrm{cv}} \\
\mathbf{C}_{\mathrm{ac}} & \mathbf{C}_{\mathrm{aa}}&\mathbf{C}_{\mathrm{av}}\\
\mathbf{C}_{\mathrm{vc}} & \mathbf{C}_{\mathrm{va}}&\mathbf{C}_{\mathrm{vv}}
\end{pmatrix},\label{pFLMO2CMO}
\end{eqnarray}
the Fock matrix can be re-block-diagonalized by the unitary transformation\cite{OpenFLMO}
\begin{eqnarray}
\mathbf{U}&=&\mathbf{C}\mathbf{T},\label{Umat}\\
\mathbf{T}&=&
\begin{pmatrix}
\mathbf{C}_{\mathrm{cc}}^\dag(\mathbf{C}_{\mathrm{cc}}\mathbf{C}_{\mathrm{cc}}^\dag)^{-1/2}&\mathbf{0}&\mathbf{0}\\
\mathbf{0}&\mathbf{C}_{\mathrm{aa}}^\dag(\mathbf{C}_{\mathrm{aa}}\mathbf{C}_{\mathrm{aa}}^\dag)^{-1/2}&\mathbf{0}\\
\mathbf{0}&\mathbf{0}&\mathbf{C}_{\mathrm{vv}}^\dag(\mathbf{C}_{\mathrm{vv}}\mathbf{C}_{\mathrm{vv}}^\dag)^{-1/2}\end{pmatrix},\label{Tmat}
\end{eqnarray}
where the unitary matrix $\mathbf{T}$ is just the mapping between the CMOs $\{\psi_{\mathrm{c}}^{\mathrm{CMO}},\psi_{\mathrm{a}}^{\mathrm{CMO}},\psi_{\mathrm{v}}^{\mathrm{CMO}}\}$ and the LMOs $\{\psi_{\mathrm{c}}^{\mathrm{LMO}},\psi_{\mathrm{a}}^{\mathrm{LMO}},\psi_{\mathrm{v}}^{\mathrm{LMO}}\}$, viz.,
\begin{eqnarray}
(\psi_{\mathrm{c}}^{\mathrm{LMO}}, \psi_{\mathrm{a}}^{\mathrm{LMO}}, \psi_{\mathrm{v}}^{\mathrm{LMO}}) =(\phi_{\mathrm{c}}^{\mathrm{pFLMO}}, \phi_{\mathrm{a}}^{\mathrm{pFLMO}}, \phi_{\mathrm{v}}^{\mathrm{pFLMO}}) \mathbf{U}
=(\psi_{\mathrm{c}}^{\mathrm{CMO}}, \psi_{\mathrm{a}}^{\mathrm{CMO}}, \psi_{\mathrm{v}}^{\mathrm{CMO}}) \mathbf{T}.\label{CMO2LMO}
\end{eqnarray}
It can be proven\cite{ACR-FLMO} that the so-obtained LMOs change least (in the least-squares sense) from the pFLMOs and can therefore be termed FLMOs\cite{F2M}.
The very key for the success of F2M is that the pFLMOs are local not only in space by construction but also in energy in the sense that
their occupation numbers do not deviate significantly from the ideal value 2, 1 or 0.
Different partitions of the same molecule may affect the locality of the FLMOs but only to a minor extent,
which is tolerable given the dramatic gain in efficiency. Note in passing that, at variance with this noniterative ``top-down least-change''
(``diagonalize-then-block-diagonalize'') algorithm, an iterative  ``bottom-up least-change'' algorithm is also available, which does not invoke the global CMOs at all\cite{OpenFLMO}.

Now suppose that the low-lying states of a system lies in the space spanned by a set of valence AOs (which requires minimal chemical/physical intuitions). It is obvious that
some core AOs can also be chosen here if core excitations are to be investigated. Such occupied and unoccupied AOs
can readily be taken from atomic NO type of generally contracted basis or from ROHF calculations of spherical, unpolarized atomic configurations.
The occupied AOs are further classified into doubly and singly occupied ones. The number of the latter is just that of singly occupied molecular orbitals (MO).
The doubly occupied AOs are first symmetrically orthonormalized, leading to $\{\phi_p^{\mathrm{OAO}}\}_{p=1}^{K_\mathrm{c}}$. After projecting out $\{\phi_p^{\mathrm{OAO}}\}_{p=1}^{K_\mathrm{c}}$,
the singly occupied AOs
are then symmetrically orthonormalized, leading to $\{\phi_p^{\mathrm{OAO}}\}_{p=K_c+1}^{K_\mathrm{o}}$ ($K_{\mathrm{o}}=K_\mathrm{c}+K_\mathrm{a}$). After projecting out both $\{\phi_p^{\mathrm{OAO}}\}_{p=1}^{K_\mathrm{c}}$
and $\{\phi_p^{\mathrm{OAO}}\}_{p=K_\mathrm{c}+1}^{K_{\mathrm{o}}}$,
the unoccupied AOs are finally symmetrically orthonormalized, leading to $\{\phi_p^{\mathrm{OAO}}\}_{p=K_\mathrm{c}+K_\mathrm{a}+1}^{K}$ ($K=K_\mathrm{o}+K_\mathrm{v}$).
Such orthonormal AOs (OAO) can generally be written as
\begin{equation}
\phi_p^{\mathrm{OAO}}=\sum_{\mu}^K\tilde{\chi}_{\mu}\tilde{A}_{\mu p},\quad p\in [1, K],
\end{equation}
where the primitive functions $\{\tilde{\chi}_{\mu}\}$ need not be the same as $\{\chi_{\mu}\}$ used in the molecular calculation.
Projecting the converged molecular Fock operator
\begin{equation}
f=\sum_{\mu\nu}|\chi_{\mu}\rangle(\mathbf{S}_{11}^{-1}\mathbf{F}\mathbf{S}_{11}^{-1})_{\mu\nu}\langle\chi_{\nu}|,\quad (\mathbf{S}_{11})_{\mu\nu}=\langle\chi_{\mu}|\chi_{\nu}\rangle
\end{equation}
 onto the OAO basis leads to
\begin{eqnarray}
\tilde{\mathbf{F}}&=&\mathbf{S}_{21}\mathbf{S}_{11}^{-1}\mathbf{F}\mathbf{S}_{11}^{-1}\mathbf{S}_{12},\quad (S_{12})_{\mu\nu}=\langle\chi_{\mu}|\tilde{\chi}_{\nu}\rangle=(S_{21})_{\nu\mu}^*.\label{Foao}
\end{eqnarray}
Since the so-defined OAOs play exactly the same role as the pFLMOs, the $\tilde{\mathbf{F}}$ matrix \eqref{Foao} in the form of Eq. \eqref{F:CMO3}
 can be block-diagonalized via the unitary transformation $\tilde{\mathbf{U}}$ in the form of Eq. \eqref{Umat}, leading to
a set of pre-LMOs,
\begin{eqnarray}
\tilde{\psi}_p^{\mathrm{LMO}}=\sum_q^K\phi_q^{\mathrm{OAO}} \tilde{U}_{qp}=\sum_{\mu}^K\tilde{\chi}_{\mu}\tilde{C}^{\mathrm{LMO}}_{\mu p}, \quad p\in [1, K]; \quad \tilde{\mathbf{C}}^{\mathrm{LMO}}=\tilde{\mathbf{A}}\tilde{\mathbf{U}}.\label{pre-LMO}
\end{eqnarray}
The corresponding pre-CMOs read
\begin{eqnarray}
\tilde{\psi}_p^{\mathrm{CMO}}=\sum_{\mu}^K\tilde{\chi}_{\mu}\tilde{C}^{\mathrm{CMO}}_{\mu p}, \quad p\in [1, K]; \quad \tilde{\mathbf{C}}^{\mathrm{CMO}}=\tilde{\mathbf{C}}^{\mathrm{LMO}}\tilde{\mathbf{T}}^\dag,\label{pre-CMO}
\end{eqnarray}
where $\tilde{\mathbf{T}}$ in the form of Eq. \eqref{Tmat} is one part of $\tilde{\mathbf{U}}$.
Note in passing that, given the simplicity of the chosen AOs, the pre-LMOs \eqref{pre-LMO} can also be obtained by a top-down localization\cite{ACR-FLMO} of the pre-CMOs \eqref{pre-CMO} via, e.g., the Pipek and Mezey (PM) scheme\cite{LMO-PM}. In this case, the OAOs can be obtained simply by a one-step symmetric orthonormalization of all the AOs,
without the need of classifying their occupations. Instead, the singly occupied pre-CMOs can readily
be determined before the top-down localization is executed, since they should overlap most with the singly occupied molecular CMOs
(cf. Eqs. \eqref{Bmat} and \eqref{Nocc}).

To identify the active orbitals, we first calculate the absolute overlaps between the pre-LMOs and molecular ROHF LMOs of the same occupancy
\begin{eqnarray}
(B_{\mathrm{x}}^{\mathrm{Y}})_{pi}=|\langle\tilde{\psi}_{\mathrm{x},p}^{\mathrm{Y}}|\psi_{\mathrm{x},i}^{\mathrm{Y}}\rangle|
=|(\tilde{\mathbf{C}}^{\mathrm{y}\dag}_{\mathrm{x}}\mathbf{S}_{21}\mathbf{C}^{\mathrm{y}}_{\mathrm{x}})_{pi}|,
\quad \mathrm{x}=\mathrm{a}, \mathrm{c}, \mathrm{v}; \quad \mathrm{Y}=\mathrm{LMO}, \mathrm{CMO},\label{Bmat}
\end{eqnarray}
where $\mathbf{C}^{\mathrm{LMO}}$ is the $(N_{\mathrm{c}}+N_{\mathrm{a}}+N_{\mathrm{v}})$-dimensional coefficient matrix of the molecular ROHF LMOs in the basis $\{\chi_{\mu}\}$ and $\mathbf{B}_{\mathrm{x}}^{\mathrm{LMO}}$ is a $K_{\mathrm{x}}\times N_{\mathrm{x}}$ rectangular matrix.
For each row of $\mathbf{B}_{\mathrm{x}}^{\mathrm{LMO}}$, one can identify the largest matrix element so as to form a closed
$K_{\mathrm{x}}\times K_{\mathrm{x}}$ subblock. In particular, $B_{\mathrm{a}}^{\mathrm{LMO}}$ should be a strongly diagonal $K_{\mathrm{a}}\times K_{\mathrm{a}}$ matrix,
with all diagonal elements close to one. Otherwise, the occupations of the pre-CMOs must be reassigned until this criterion is fulfilled.
As a matter of fact, for most organic systems, the $K_{\mathrm{c}}\times K_{\mathrm{c}}$ subblock of $B_{\mathrm{c}}^{\mathrm{LMO}}$ and
the $K_{\mathrm{v}}\times K_{\mathrm{v}}$ subblock of $B_{\mathrm{v}}^{\mathrm{LMO}}$ are also strongly diagonal, as can be seen from
Tables \ref{OverlapOccLMO} and \ref{OverlapVirLMO} for
the case of \ce{(C4SH3)-(CH2)10-(C4SH3)} (see Ref. \citenum{DLPNO-NEVPT2} for the geometry).
Actually, following the maximum occupation method (mom) for excited solutions of the HF equation\cite{Triad},
an occupation number $n_i$ can be defined for each molecular LMO $\psi_i^{\mathrm{LMO}}$ in the space spanned by the pre-LMOs,
\begin{eqnarray}
n_{\mathrm{x},i}&=&\langle\psi_{\mathrm{x},i}^{\mathrm{Y}}|P_{\mathrm{x}}^{\mathrm{Y}}|\psi_{\mathrm{x},i}^{\mathrm{Y}}\rangle
=\sum_{p\in\mathrm{x}}^{K_{\mathrm{x}}}[(B_{\mathrm{x}}^{\mathrm{Y}})_{pi}]^2,\quad \mathrm{x}=\mathrm{a}, \mathrm{c}, \mathrm{v}; \quad \mathrm{Y}=\mathrm{LMO}, \mathrm{CMO},\label{Nocc}\\
P_{\mathrm{x}}^{\mathrm{Y}}&=&\sum_{p\in\mathrm{x}}^{K_{\mathrm{x}}}|\tilde{\psi}_{\mathrm{x},p}^{\mathrm{Y}}\rangle\langle\tilde{\psi}_{\mathrm{x},p}^{\mathrm{Y}}|,
\end{eqnarray}
such that the $K_{\mathrm{c}}$/$K_{\mathrm{a}}$/$K_{\mathrm{v}}$ doubly/singly/zero occupied active orbitals are just those with
the largest occupation numbers, see again
Tables \ref{OverlapOccLMO} and \ref{OverlapVirLMO}.

If CMOs instead of LMOs are to be used in CASSCF calculations, the corresponding $\mathbf{B}_{\mathrm{x}}^{\mathrm{CMO}}$ matrices \eqref{Bmat} are
no longer block-diagonal because of the relation $\mathbf{B}^{\mathrm{CMO}}=\tilde{\mathbf{T}}\mathbf{B}^{\mathrm{LMO}}\mathbf{T}^\dag$, i.e.,
the block-diagonal matrix $\mathbf{B}^{\mathrm{LMO}}$ is transformed generally to a full matrix $\mathbf{B}^{\mathrm{CMO}}$,
see Tables \ref{OverlapOccCMO} and \ref{OverlapVirCMO} for the showcase of \ce{(C4SH3)-(CH2)10-(C4SH3)}.
In this case, we first carry out a singular value decomposition (SVD) of $\mathbf{B}_{\mathrm{x}}^{\mathrm{CMO}}$
\begin{eqnarray}
\mathbf{B}_{\mathrm{x}}^{\mathrm{CMO}}=\mathbf{L}_{\mathrm{x}}^{\mathrm{CMO}}\lambda^{\mathrm{CMO}}_{\mathrm{x}}\mathbf{R}_{\mathrm{x}}^{\mathrm{CMO}\dag},\quad \mathrm{x}=\mathrm{a}, \mathrm{c}, \mathrm{v}.
\end{eqnarray}
The right vectors $\mathbf{R}_{\mathrm{x},i}^{\mathrm{CMO}}$ can then be employed to transform the molecular CMOs\cite{ACR-FLMO}
\begin{eqnarray}
\bar{\psi}_{\mathrm{x},i}^{\mathrm{CMO}}=\sum_{j\in\mathrm{x}}^{N_{\mathrm{x}}}\psi_{\mathrm{x},j}^{\mathrm{CMO}} R_{\mathrm{x},ji}^{\mathrm{CMO}}, \quad \mathrm{x}=\mathrm{a}, \mathrm{c}, \mathrm{v}; \quad i\in [1, N_{\mathrm{x}}].
\end{eqnarray}
Since the number of nonzero singular values $\lambda^{\mathrm{CMO}}_{\mathrm{x},i}$ is just $K_{\mathrm{x}}$, the rotated CMOs are clearly separated into two groups: those with nonzero $\lambda^{\mathrm{CMO}}_{\mathrm{x},i}$
are just the active orbitals whereas those with zero $\lambda^{\mathrm{CMO}}_{\mathrm{x},i}$ are the inactive ones (doubly or zero occupied), which can further be semi-canonicalized.

\begin{figure}
\centering
\includegraphics[width=0.6\textwidth]{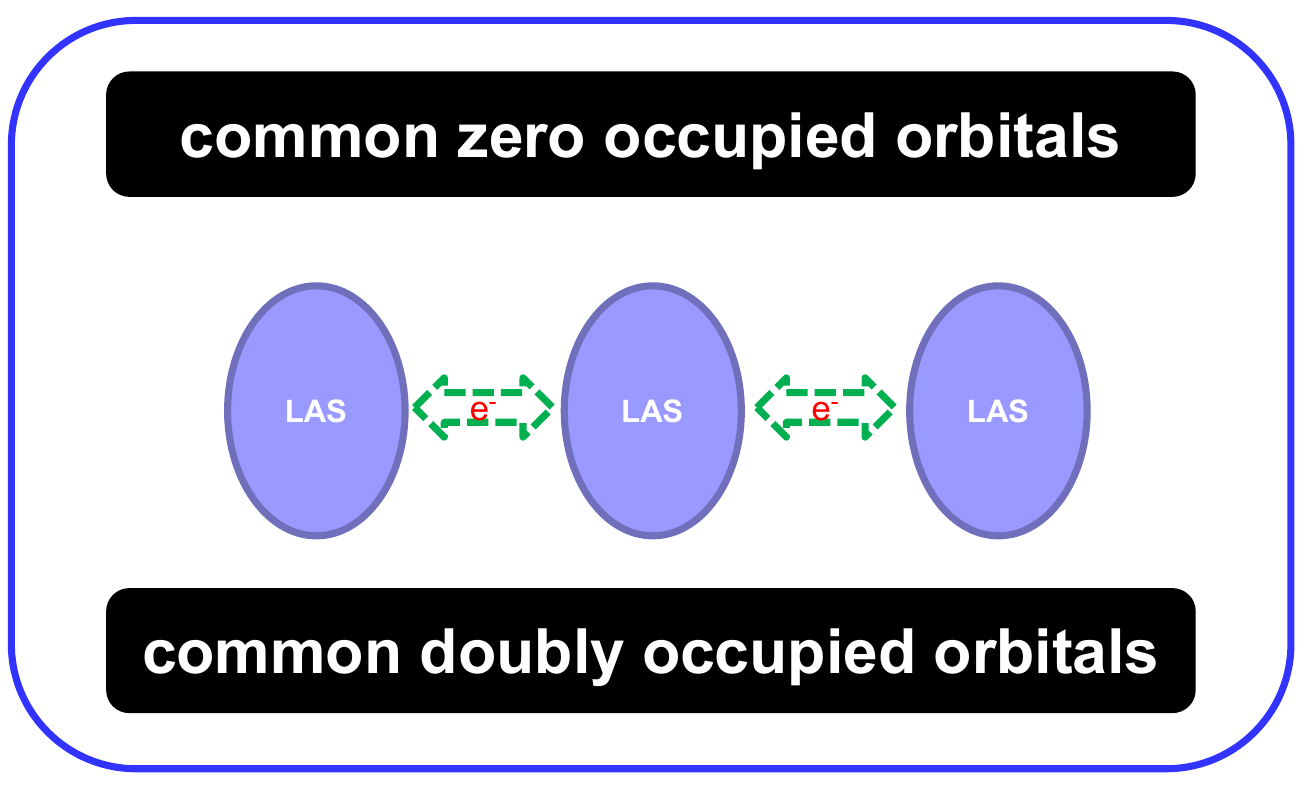}
\caption{Illustration of local active spaces (LAS) in terms of LMOs.}
\label{LAS}
\end{figure}

The above automated selection of active orbitals is clearly different from the atomic valence active space (AVAS) approach\cite{AVAS}, where the $K$ valence AOs are used to select both the occupied and unoccupied active orbitals,
thereby leading to in total $2K$ active orbitals. The number of rotated active orbitals can be reduced therein by discarding those with small singular values but then there is no guarantee to
maintain the same CAS for all geometries of complex systems.
In contrast, the $K$ AOs (which may include core AOs) are separated here into doubly, singly and zero occupied ones, which are then used as probes to select
the corresponding active orbitals.
The number of the so-selected active orbitals is hence just $K$ instead of $2K$. This is precisely the spirit of F2M\cite{F2M}. As a matter of fact, some pFLMOs from
subsystems calculations\cite{ACR-FLMO} can also be used to form the (global) pre-LMOs. This way, the task of selecting active orbitals for a large system is split into selections of
subsystem active orbitals. If wanted, the number of active orbitals can be reduced by first semicanonicalizing those pre-LMOs of a fragment (e.g., an aromatic ring; cf. \eqref{Foao}) and then picking up
those of right ``energies'' (NB: semicanonicalization means here that the doubly, singly and zero occupied pre-LMOs
are canonicalized separately). Since such regional pre-CMOs are still localized on the chosen fragment, they can be used simply as pre-LMOs, provided that the molecular LMOs localized on the same fragment
are also semicanonicalized in the same way. The CAS will remain the same for all geometries as long as the chosen fragments maintain their characters. A similar idea was also adopted by the PiOS apprach\cite{PiOS} but only for $\pi$-orbital spaces (PiOS)
of aromatic rings, as implied by the acronym. Alternatively, one can define several local active spaces (LAS)\cite{LAS-DMET} and further impose certain restrictions on the excitations between the LASs, as illustrated in Fig. \ref{LAS}.

\newpage

\begin{threeparttable}
  \caption{Absolute overlaps between the six occupied pre-LMOs and all occupied molecular HF-LMOs for \ce{(C4SH3)-(CH2)10-(C4SH3)}}
  \tabcolsep=10pt
  \scriptsize
  \begin{tabular}[t]{@{}ccccccccccccc@{}}
   \hline\hline
   \diagbox[width=10em,height=3em,trim=l]{pre-LMO\tnote{a}}{active LMO}      & 1     & 2     & 3     & 4     & 5     & 6     & $\cdots$\tnote{b} \\
    \hline
1	&	\textbf{0.996} 	&	0.000 	&	0.000 	&	0.000 	&	0.000 	&	0.032 	&	$\cdots$\\
2	&	0.000 	&	\textbf{0.996} 	&	0.000 	&	0.000 	&	0.032 	&	0.000 	&	$\cdots$\\
3	&	0.032 	&	0.000 	&	\textbf{0.992} 	&	0.032 	&	0.032 	&	0.000 	&	$\cdots$\\
4	&	0.000 	&	0.000 	&	0.000 	&	\textbf{0.992} 	&	0.000 	&	0.032 	&	$\cdots$\\
5	&	0.000 	&	0.000 	&	0.000 	&	0.000 	&	\textbf{0.986} 	&	0.000 	&	$\cdots$\\
6	&	0.000 	&	0.032 	&	0.032 	&	0.000 	&	0.000 	&	\textbf{0.986} 	&	$\cdots$\\
          & 0.992\tnote{c}  & 0.992\tnote{c}  & 0.985\tnote{c}  & 0.985\tnote{c}  & 0.974\tnote{c}  & 0.974\tnote{c}  & $\cdots$\\
   \hline \hline
  \end{tabular}\label{OverlapOccLMO}
  \begin{tablenotes}
  \item [a] Five $n\textrm{p}_z$ (n = 2 for C and 3 for S) orbitals from each of the terminal 5-membered rings \ce{C4S}.
  \item [b] Columns (core orbitals) with values smaller than 0.009.
  \item [c] Occupation number.
  \end{tablenotes}
\end{threeparttable}

\newpage

\begin{threeparttable}
  \caption{Absolute overlaps between the four unoccupied pre-LMOs and all unoccupied molecular HF-LMOs for \ce{(C4SH3)-(CH2)10-(C4SH3)}}
  \tabcolsep=20pt
  \scriptsize
  \begin{tabular}[t]{@{}ccccccccccccc@{}}
   \hline\hline
   \diagbox[width=10em,height=3em,trim=l]{pre-LMO\tnote{a}}{active LMO}  & 1  & 2  & 3  & 4  & $\cdots$\tnote{b} \\
    \hline
1	&	\textbf{0.894} 	&	0.000 	&	0.032 	&	0.000 	&	$\cdots$\\
2	&	0.000 	&	\textbf{0.894} 	&	0.000 	&	0.032 	&	$\cdots$\\
3	&	0.032 	&	0.000 	&	\textbf{0.864} 	&	0.000 	&	$\cdots$\\
4	&	0.000 	&	0.032 	&	0.000 	&	\textbf{0.864} 	&	$\cdots$\\
          & 0.801\tnote{c}  & 0.801\tnote{c}  & 0.748\tnote{c}  & 0.748\tnote{c}  & $\cdots$\\
   \hline \hline
  \end{tabular}\label{OverlapVirLMO}
  \begin{tablenotes}
  \item [a] Five $n\textrm{p}_z$ (n = 2 for C and 3 for S) orbitals from each of the terminal 5-membered rings \ce{C4S}.
  \item [b] Columns (virtual orbitals) with values smaller than 0.060.
  \item [c] Occupation number.
  \end{tablenotes}
\end{threeparttable}

\newpage

\begin{threeparttable}
  \caption{Absolute overlaps between the six occupied pre-CMOs and all occupied molecular HF-CMOs for \ce{(C4SH3)-(CH2)10-(C4SH3)}}
  \tabcolsep=4pt
  \scriptsize
  \begin{tabular}[t]{@{}ccccccccccccc@{}}
   \hline\hline
   \diagbox[width=10em,height=3em,trim=l]{pre-CMO\tnote{a}}{active CMO} & 1 & 2 & 3 & 4 & 5 & 6 & 7 & 8 & 9 & 10 & $\cdots$\tnote{b} \\
    \hline
1	&	0.032 	&	0.000 	&	0.055 	&	0.000 	&	0.543 	&	0.506  	&	0.404 	&	0.305 	&	0.279 	&	0.205 	&	$\cdots$\\
2	&	0.000 	&	0.032 	&	0.000 	&	0.055 	&	0.539 	&	0.510  	&	0.407 	&	0.302 	&	0.277 	&	0.205 	&	$\cdots$\\
3	&	0.809 	&	0.581 	&	0.032 	&	0.000 	&	0.000 	&	0.000  	&	0.000 	&	0.000 	&	0.000 	&	0.000 	&	$\cdots$\\
4	&	0.581 	&	0.809 	&	0.000 	&	0.032 	&	0.000 	&	0.000  	&	0.000 	&	0.000 	&	0.000 	&	0.000 	&	$\cdots$\\
5	&	0.032 	&	0.000 	&	0.931 	&	0.173 	&	0.063 	&	0.134  	&	0.045 	&	0.134 	&	0.100 	&	0.118 	&	$\cdots$\\
6	&	0.000 	&	0.032 	&	0.173 	&	0.931 	&	0.063 	&	0.134  	&	0.045 	&	0.134 	&	0.100 	&	0.118 	&	$\cdots$\\
      & 0.993\tnote{c}  & 0.993\tnote{c}  & 0.900\tnote{c}  & 0.900\tnote{c}  & 0.593\tnote{c}  & 0.552\tnote{c}  & 0.333\tnote{c}  & 0.220\tnote{c}  & 0.175\tnote{c}  & 0.112\tnote{c}  & $\cdots$\\
   \hline \hline
  \end{tabular}\label{OverlapOccCMO}
  \begin{tablenotes}
  \item [a] Five $n\textrm{p}_z$ (n = 2 for C and 3 for S) orbitals from each of the terminal 5-membered rings \ce{C4S}.
  \item [b] Columns (core orbitals) with values smaller than 0.025.
  \item [c] Occupation number.
  \end{tablenotes}
\end{threeparttable}

\newpage

\begin{threeparttable}
  \caption{Absolute overlaps between the four unoccupied pre-CMOs and all unoccupied molecular HF-CMOs for \ce{(C4SH3)-(CH2)10-(C4SH3)}}
  \tabcolsep=5pt
  \scriptsize
  \begin{tabular}[t]{@{}ccccccccccccc@{}}
   \hline\hline
   \diagbox[width=10em,height=3em,trim=l]{pre-CMO\tnote{a}}{active CMO}  & 1  & 2  & 3  & 4  & 5  & 6 & 7 & 8 & 9 & $\cdots$\tnote{b} \\
    \hline
1	&	0.709 	&	0.608 	&	0.032 	&	0.105 	&	0.045 	&	0.126 	&	0.000 	&	0.032 	&	0.055 	&	$\cdots$\\
2	&	0.608 	&	0.709 	&	0.032 	&	0.105 	&	0.045 	&	0.126 	&	0.000 	&	0.032 	&	0.055 	&	$\cdots$\\
3	&	0.089 	&	0.077 	&	0.586 	&	0.559 	&	0.344 	&	0.261 	&	0.161 	&	0.158 	&	0.145 	&	$\cdots$\\
4	&	0.077 	&	0.089 	&	0.586 	&	0.560 	&	0.344 	&	0.261 	&	0.161 	&	0.158 	&	0.145 	&	$\cdots$\\
      & 0.886\tnote{c}  & 0.886\tnote{c}  & 0.688\tnote{c}  & 0.649\tnote{c}  & 0.240\tnote{c}  & 0.168\tnote{c}  & 0.052\tnote{c}  & 0.052\tnote{c}  & 0.048\tnote{c}   & $\cdots$\\
   \hline \hline
  \end{tabular}\label{OverlapVirCMO}
  \begin{tablenotes}
  \item [a] Five $n\textrm{p}_z$ (n = 2 for C and 3 for S) orbitals from each of the terminal 5-membered rings \ce{C4S}.
  \item [b] Columns (virtual orbitals) with values smaller than 0.011.
  \item [c] Occupation number.
  \end{tablenotes}
\end{threeparttable}

\subsection{Subspace Matching}\label{lCAS}
Given the guess orbitals $\{\psi_{p}^{\mathrm{Y}(0)}\}_{p=1}^{N_{\mathrm{orb}}}$ ($\mathrm{Y}=\mathrm{LMO},\mathrm{CMO}$), we are ready to perform the CASSCF calculation.
Noticing that each iteration $k$ will generate delocalized CMOs $\{\psi_{i}^{\mathrm{CMO}(k)}\}_{i=1}^{N_{\mathrm{orb}}}$ and
a proper reassignment of the core, active and virtual spaces has to be performed, we first calculate the occupation numbers of all $\{\psi_{i}^{\mathrm{CMO}(k)}\}_{i=1}^{N_{\mathrm{orb}}}$
in the core space $\{\psi_p^{\mathrm{Y}(k-1)}\}_{i=1}^{N_\mathrm{c}}$ of the previous iteration
(cf. Eq. \eqref{Nocc}), so as to identify the core orbitals $\{\psi_i^{\mathrm{CMO}(k)}\}_{p=1}^{N_\mathrm{c}}$
of the present iteration with the largest $N_{\mathrm{c}}$ occupation numbers $\{n_{\mathrm{c},i}\}_{i=1}^{N_{\mathrm{c}}}$.
Then, the occupation numbers  of
the remaining $\{\psi_{i}^{\mathrm{CMO}(k)}\}_{i=N_\mathrm{c}+1}^{N_{\mathrm{orb}}}$ in the active space $\{\psi_p^{\mathrm{Y}(k-1)}\}_{p=N_{\mathrm{c}}+1}^{N_{\mathrm{c}}+K}$
are calculated to identify the $K$ active orbitals, again with the largest occupation numbers $\{n_{\mathrm{a},i}\}_{i=1}^{K}$. The left are obviously virtual orbitals.
In case that some $\{n_{\mathrm{a},j}\}$ are very close to the smallest of $\{n_{\mathrm{a},i}\}_{i=1}^{K}$, it would indicate the prechosen AOs/pFLMOs are not enough
and should hence be expanded. As a matter of fact, even in this case, the calculation need not be terminated but
can still proceed ahead by just putting such orbitals into the active space,
which will be stabilized in subsequent iterations.
This way, the orbitals of two adjacent iterations
are best matched. To localize the CASSCF-CMOs, they are first expanded as
\begin{eqnarray}
(\psi_{\mathrm{c}}^{\mathrm{CMO}(k)},\psi_{\mathrm{a}}^{\mathrm{CMO}(k)},\psi_{\mathrm{v}}^{\mathrm{CMO}(k)})=(\psi_{\mathrm{c}}^{\mathrm{Y}(k-1)},\psi_{\mathrm{a}}^{\mathrm{Y}(k-1)},\psi_{\mathrm{v}}^{\mathrm{Y}(k-1)})\begin{pmatrix}
\mathbf{C}_{\mathrm{cc}} & \mathbf{C}_{\mathrm{ca}}&\mathbf{C}_{\mathrm{cv}} \\
\mathbf{C}_{\mathrm{ac}} & \mathbf{C}_{\mathrm{aa}}&\mathbf{C}_{\mathrm{av}}\\
\mathbf{C}_{\mathrm{vc}} & \mathbf{C}_{\mathrm{va}}&\mathbf{C}_{\mathrm{vv}}
\end{pmatrix},
\end{eqnarray}
where $\mathbf{C}_{\mathrm{xy}}=\langle \psi_{\mathrm{x}}^{\mathrm{Y}(k-1)}|\psi_{\mathrm{y}}^{\mathrm{CMO}(k)}\rangle$ ($\mathrm{x}, \mathrm{y}=\mathrm{c},\mathrm{a},\mathrm{v}$).
The $\mathbf{T}$ matrix \eqref{Tmat} can then be constructed so as to transform $\{\psi_{i}^{\mathrm{CMO}(k)}\}_{i=1}^{N_{\mathrm{orb}}}$  to $\{\psi_{i}^{\mathrm{LMO}(k)}\}_{i=1}^{N_{\mathrm{orb}}}$
in view of the relation \eqref{CMO2LMO}. Since
the couplings between LMOs are much weaker than those between CMOs, generating CASSCF-LMOs in each iteration can typically facilitate the convergence.
As can be seen from Tables \ref{OverlapIFLMO} and \ref{OverlapIFCMO}, the converged CASSCF(12,10)-LMOs/CMOs do match the guess LMOs/CMOs for the
case of \ce{(C4SH3)-(CH2)10-(C4SH3)}.

\newpage
\begin{threeparttable}
  \caption{Absolute overlaps between the initial and final CAS(12,10)-LMOs of \ce{(C4SH3)-(CH2)10-(C4SH3)}}
  \tabcolsep=5pt
 \scriptsize
  \begin{tabular}[t]{@{}ccccccccccccc@{}}
    \hline \hline
    \diagbox[width=6em,height=3em,trim=l]{initial}{final}  & 1 & 2 & 3 & 4 & 5 & 6 & 7 & 8 & 9 & 10 \\
    \hline
    1     & \textbf{0.996}  & 0.000  & 0.008  & 0.000  & 0.003  & 0.000  & 0.000  & 0.011  & 0.000  & 0.006  \\
    2     & 0.000  & \textbf{0.996}  & 0.000  & 0.008  & 0.000  & 0.003  & 0.011  & 0.000  & 0.006  & 0.000  \\
    3     & 0.004  & 0.000  & \textbf{0.998}  & 0.000  & 0.001  & 0.000  & 0.000  & 0.003  & 0.000  & 0.014  \\
    4     & 0.000  & 0.004  & 0.000  & \textbf{0.998}  & 0.000  & 0.001  & 0.003  & 0.000  & 0.014  & 0.000  \\
    5     & 0.000  & 0.000  & 0.001  & 0.000  & \textbf{0.999}  & 0.000  & 0.000  & 0.001  & 0.000  & 0.011  \\
    6     & 0.000  & 0.000  & 0.000  & 0.001  & 0.000  & \textbf{0.999}  & 0.001  & 0.000  & 0.011  & 0.000  \\
    7     & 0.000  & 0.013  & 0.000  & 0.003  & 0.000  & 0.002  & \textbf{0.869}  & 0.000  & 0.015  & 0.000  \\
    8     & 0.013  & 0.000  & 0.003  & 0.000  & 0.002  & 0.000  & 0.000  & \textbf{0.869}  & 0.000  & 0.015  \\
    9     & 0.000  & 0.008  & 0.000  & 0.016  & 0.000  & 0.013  & 0.005  & 0.000  & \textbf{0.863}  & 0.000  \\
    10    & 0.008  & 0.000  & 0.016  & 0.000  & 0.013  & 0.000  & 0.000  & 0.005  & 0.000  & \textbf{0.863}  \\
    \hline \hline
  \end{tabular}\label{OverlapIFLMO}
\end{threeparttable}

\newpage
\begin{threeparttable}
  \caption{Absolute overlaps between the initial and final CAS(12,10)-CMOs of \ce{(C4SH3)-(CH2)10-(C4SH3)}}
  \tabcolsep=5pt
  \scriptsize
  \begin{tabular}[t]{@{}ccccccccccccc@{}}
    \hline \hline
    \diagbox[width=6em,height=3em,trim=l]{initial}{final} & 1 & 2 & 3 & 4 & 5 & 6 & 7 & 8 & 9 & 10 \\
    \hline
    1     & \textbf{0.997}  & 0.000  & 0.001  & 0.000  & 0.001  & 0.000  & 0.000  & 0.000  & 0.000  & 0.000  \\
    2     & 0.000  & \textbf{0.997}  & 0.001  & 0.002  & 0.000  & 0.000  & 0.000  & 0.000  & 0.000  & 0.000  \\
    3     & 0.011  & 0.001  & \textbf{0.987}  & 0.001  & 0.001  & 0.000  & 0.000  & 0.000  & 0.000  & 0.000  \\
    4     & 0.002  & 0.010  & 0.001  & \textbf{0.987}  & 0.000  & 0.001  & 0.000  & 0.000  & 0.000  & 0.000  \\
    5     & 0.003  & 0.000  & 0.005  & 0.001  & \textbf{0.999}  & 0.000  & 0.000  & 0.000  & 0.000  & 0.000  \\
    6     & 0.000  & 0.003  & 0.000  & 0.005  & 0.000  & \textbf{0.999}  & 0.000  & 0.000  & 0.000  & 0.000  \\
    7     & 0.000  & 0.000  & 0.000  & 0.000  & 0.000  & 0.000  & \textbf{0.995}  & 0.000  & 0.000  & 0.000  \\
    8     & 0.000  & 0.000  & 0.000  & 0.000  & 0.000  & 0.000  & 0.000  & \textbf{0.995}  & 0.000  & 0.000  \\
    9     & 0.000  & 0.000  & 0.000  & 0.000  & 0.000  & 0.000  & 0.005  & 0.000  & \textbf{0.991}  & 0.000  \\
    10    & 0.000  & 0.000  & 0.000  & 0.000  & 0.000  & 0.000  & 0.000  & 0.004  & 0.000  & \textbf{0.991}  \\
    \hline \hline
  \end{tabular}\label{OverlapIFCMO}
\end{threeparttable}

\newpage
\section{Illustrative examples}\label{Results}
Except for the multi-state CASPT2\cite{MSCASPT2} calculations performed with the Molcas program package\cite{OpenMolcas}, all
calculations were carried out with the BDF program package\cite{BDF1,BDF2,BDF3,BDFECC,BDFrev2020,XianCI2018}. In particular,
the Xi'an-CI module\cite{XianCI2018} in BDF was used to do the multi-state NEVPT2\cite{MSNEVPT2} and SDSPT2\cite{SDS,SDSPT2} calculations.
The valence AOs were generated by SCF calculations of spherical, unpolarized atomic configurations with the same basis functions as used for molecular calculations. Local orientations of the AOs are taken care of automatically.

\subsection{\ce{C-H} bond breaking of ethylene}\label{CHbreaking}
\begin{figure}
\centering
\includegraphics[width=0.6\textwidth]{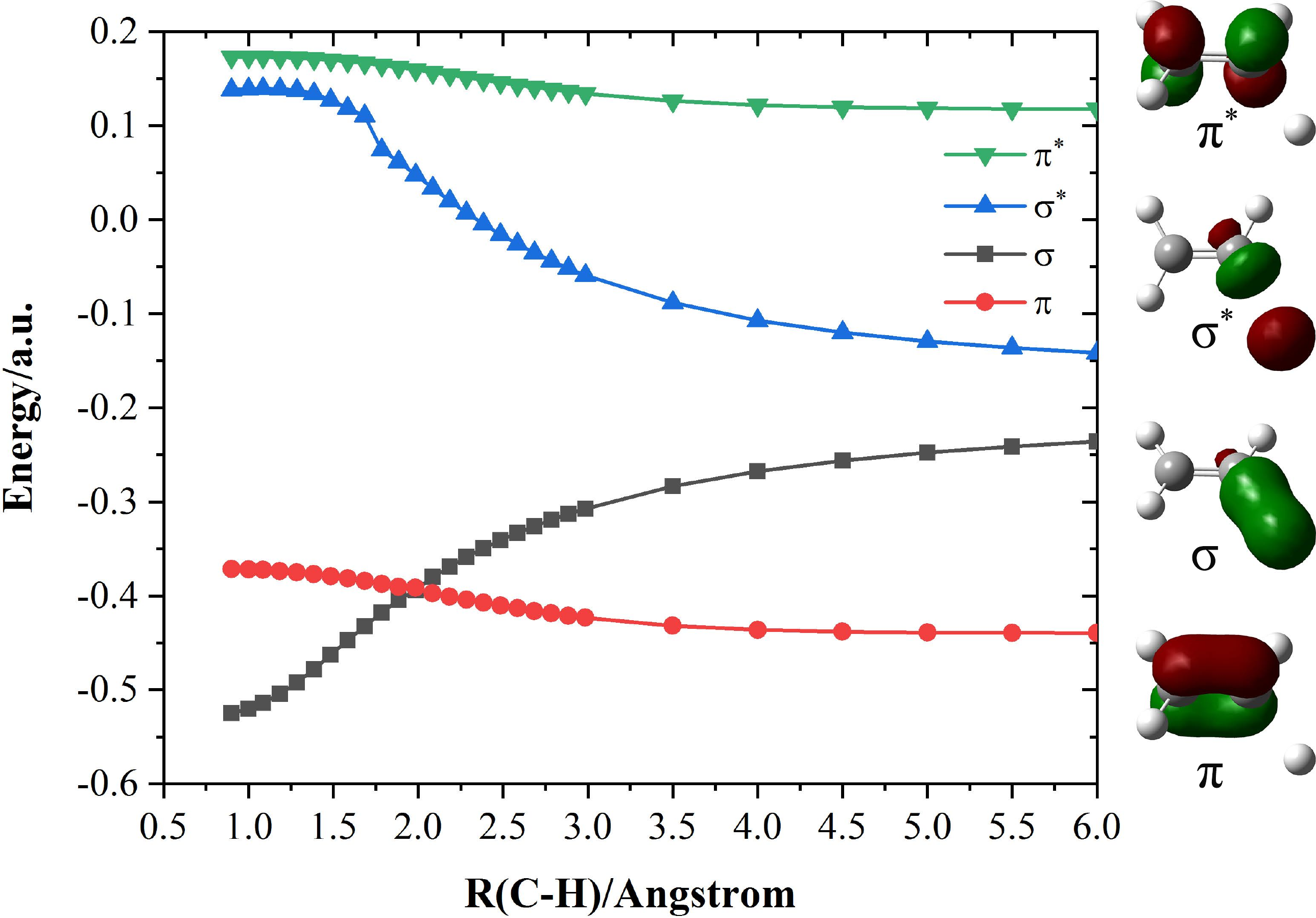}
\caption{Energies of frontier HF orbitas for the \ce{C-H} bond breaking of ethylene.}
\label{C2H4-orb}
\end{figure}

\begin{figure}
\centering
\includegraphics[width=0.6\textwidth]{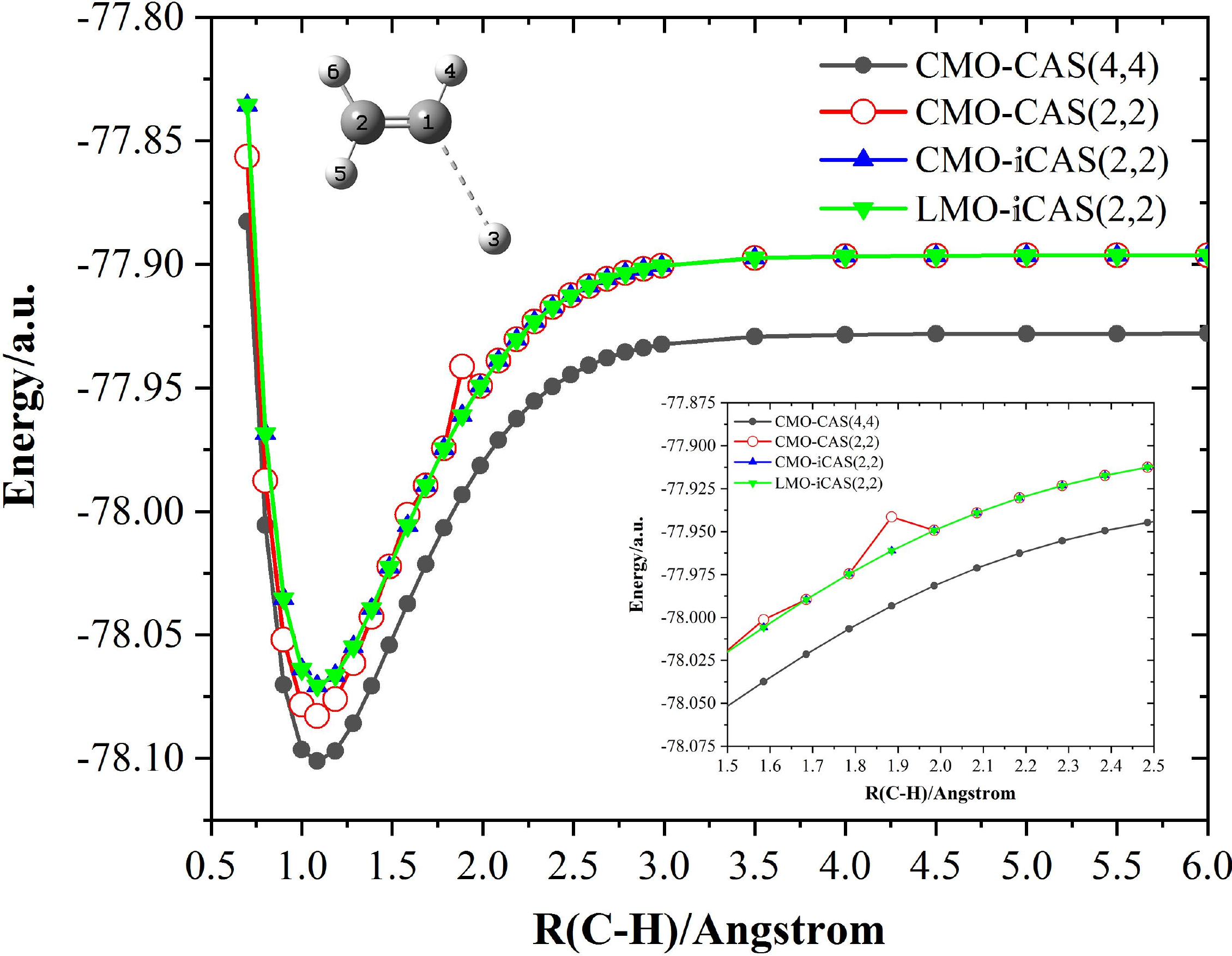}
\caption{Potential energy curves for the breaking of single \ce{C-H} bond of ethylene.}
\label{C2H4-PES-one}
\end{figure}

\begin{figure}
\centering
\includegraphics[width=0.6\textwidth]{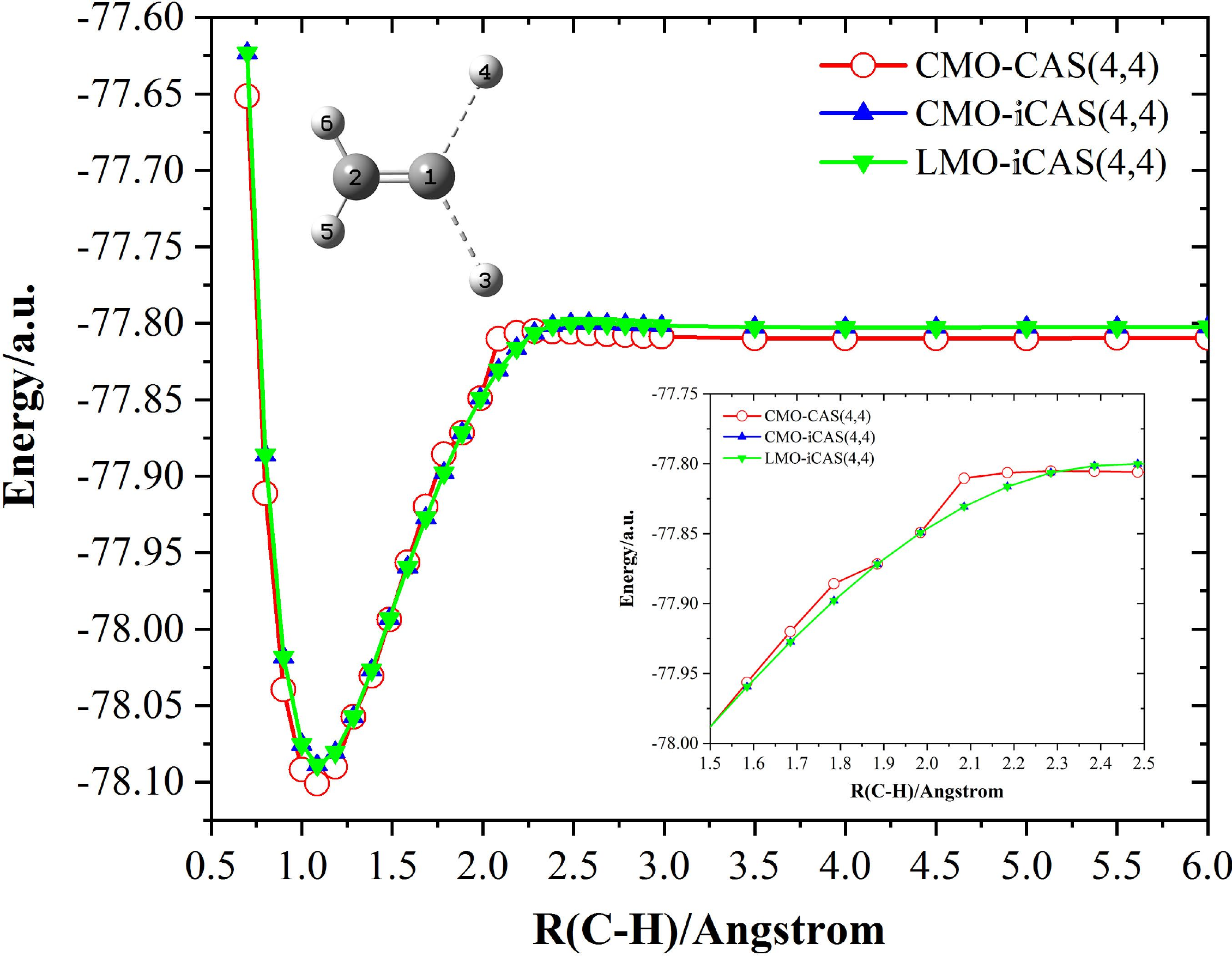}
\caption{Potential energy curves for the simultaneous breaking of the two terminal \ce{C-H} bonds of ethylene.}
\label{C2H4-PES-two}
\end{figure}

\begin{figure}
\centering
\includegraphics[width=0.6\textwidth]{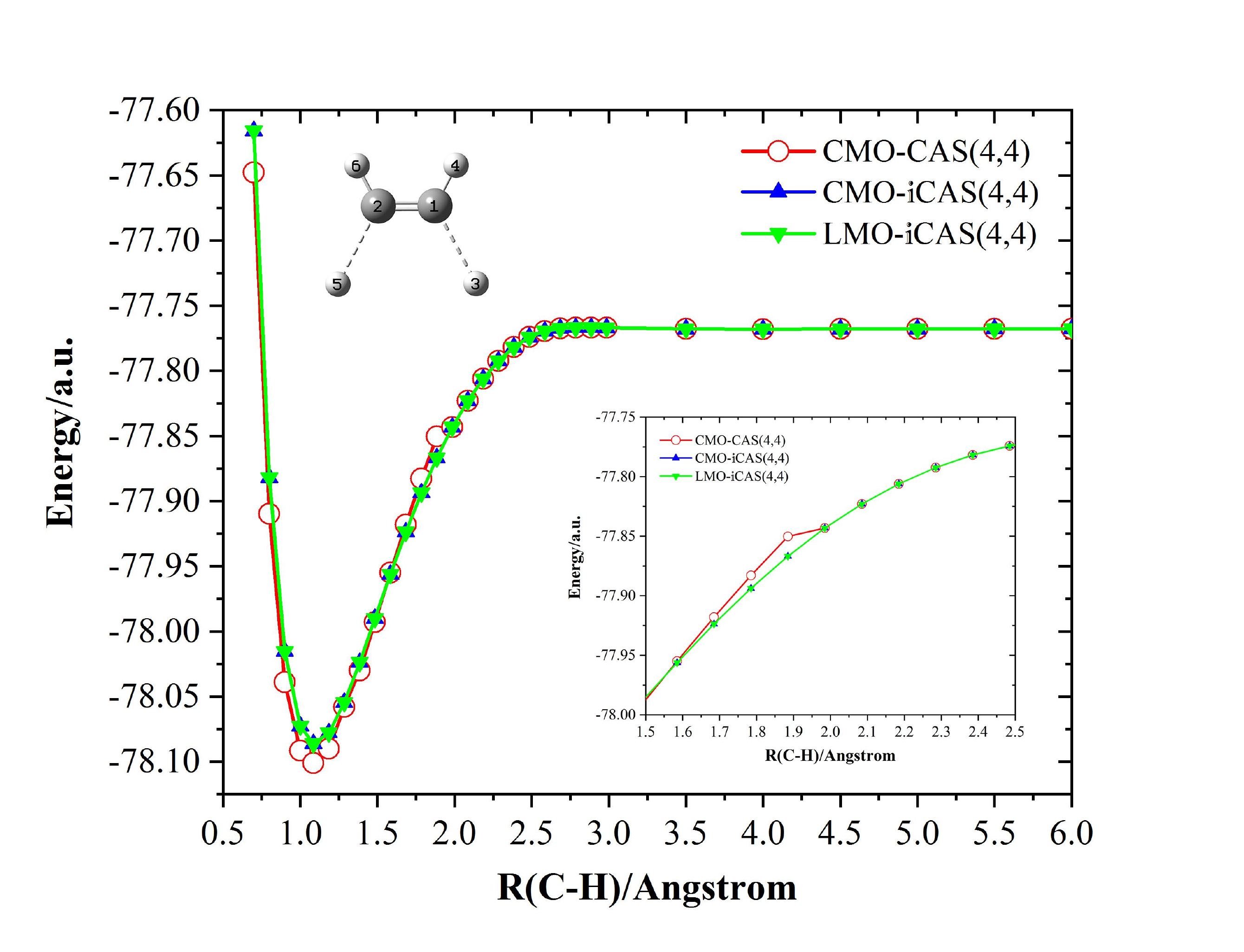}
\caption{Potential energy curves for the simultaneous breaking of the two cis \ce{C-H} bonds of ethylene.}
\label{C2H4-PES-two-cis}
\end{figure}

\begin{figure}
\centering
\includegraphics[width=0.6\textwidth]{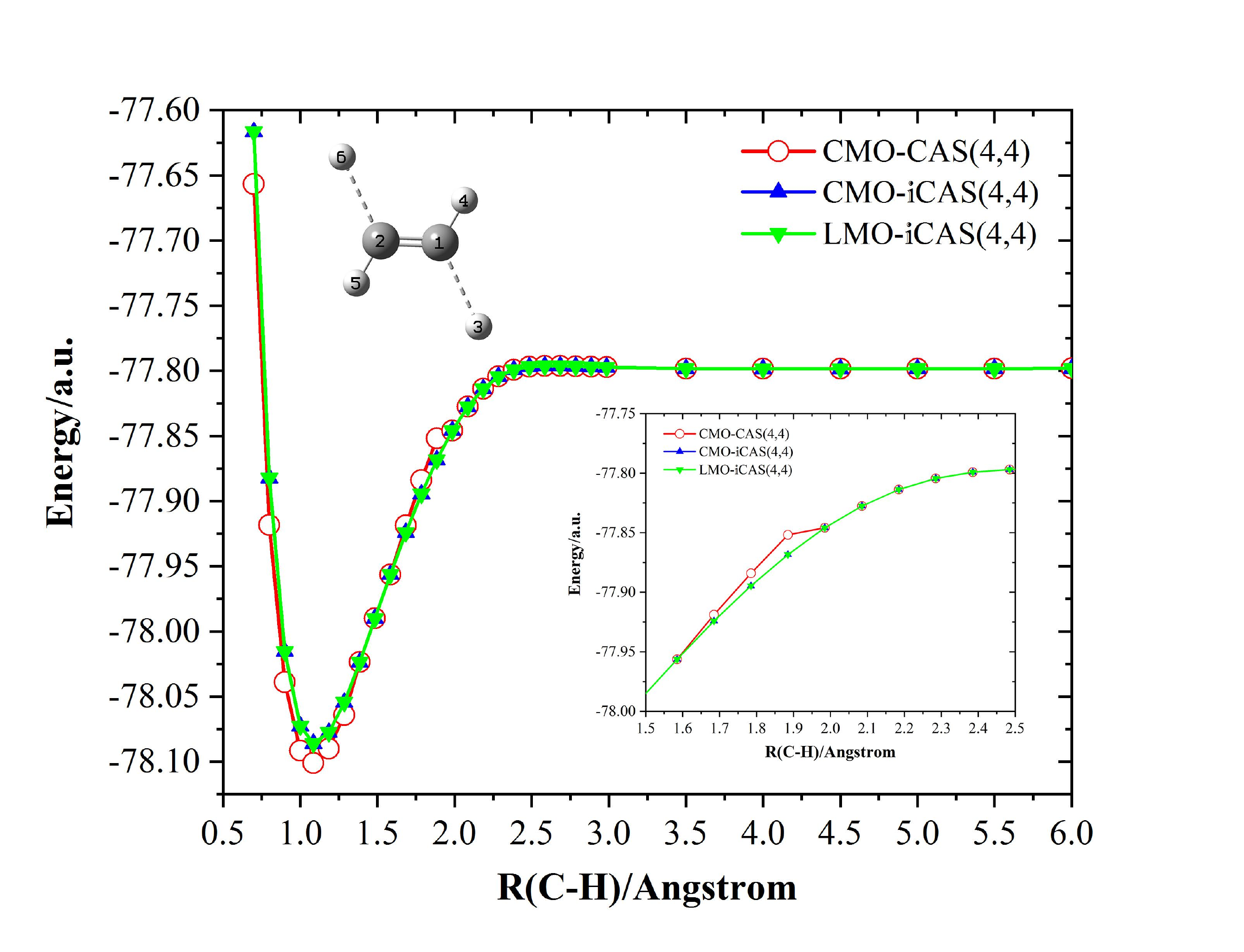}
\caption{Potential energy curves for the simultaneous breaking of the two trans \ce{C-H} bonds of ethylene.}
\label{C2H4-PES-two-trans}
\end{figure}

\begin{figure}
\centering
\includegraphics[width=0.6\textwidth]{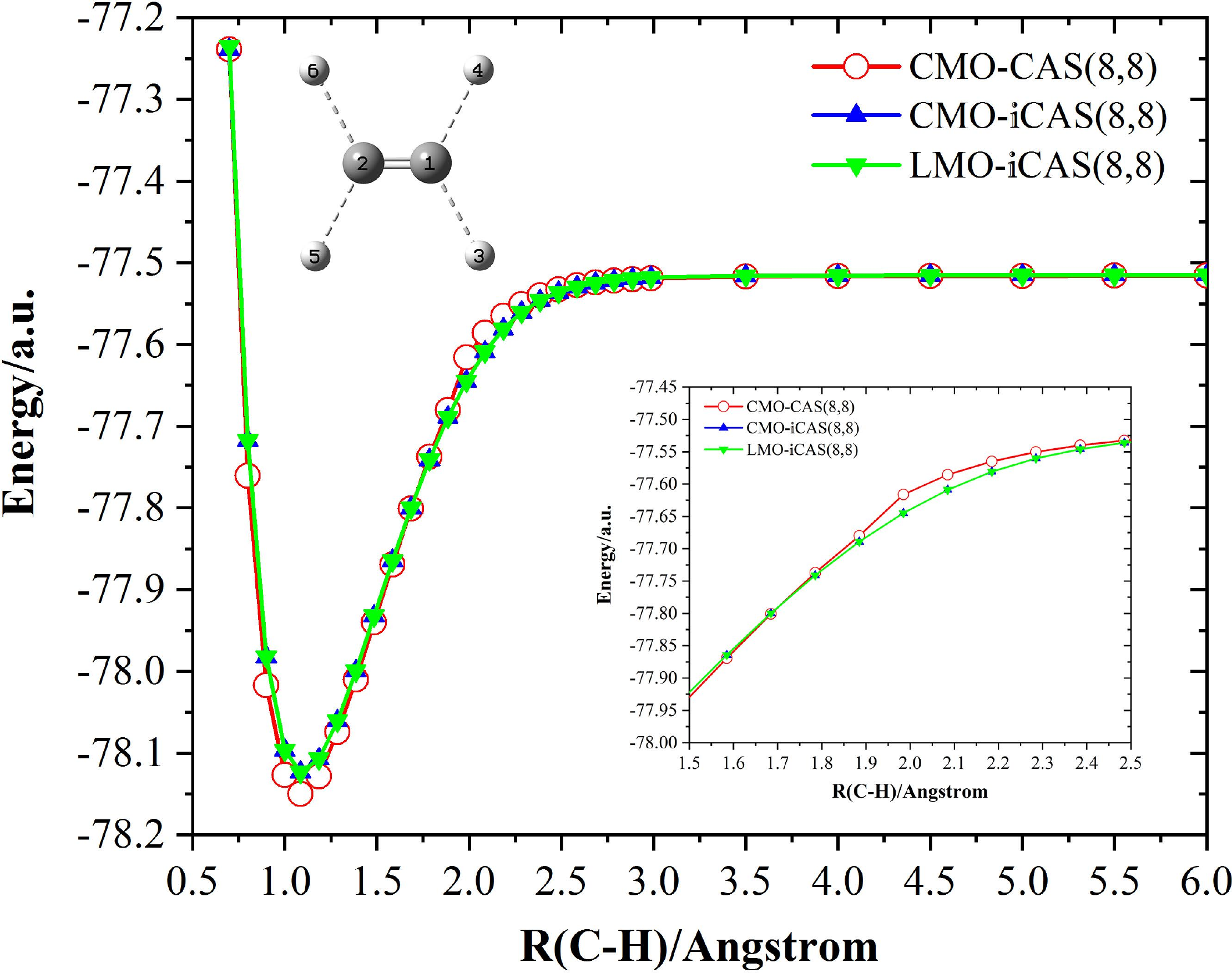}
\caption{Potential energy curves for the simultaneous breaking of four \ce{C-H} bonds of ethylene.}
\label{C2H4-PES-four}
\end{figure}

As a first application of the proposed $\mathbbm{i}$CAS approach,
we examine the breaking of one \ce{C-H} bond of ethylene, with the lengths of the other three \ce{C-H} bonds, that of the \ce{C-C} bond and the $\angle$\ce{HCH} angle set to  1.085 {\AA},  1.339 {\AA} and  124.8$^\circ$, respectively. The basis set is ANO-RCC-VDZP\cite{anorcc}.
Intuitively, the breaking of a single $\sigma$-bond requires only a minimal CAS, i.e.,
two electrons in $\sigma$ and $\sigma^\ast$. However, this picture holds only when $\sigma$ and $\sigma^\ast$ are the highest occupied (HOMO) and lowest unoccupied (LUMO) molecular orbitals, respectively, during the whole dissociation process.
As can be seen from Fig. \ref{C2H4-orb}, the $\sigma$ orbital is actually the HOMO-1 (lower than $\pi$) for distances shorter than $2.0$ {\AA} and becomes the HOMO only beyond this distance. That is,
there exists a swap of $\sigma$ and $\pi$ orbitals in the CAS(2,2) across this point. Consequently, the CMO-CAS(2,2) potential energy curve (PEC) is discontinuous nearby $2.0$ {\AA}, as can be seen from Fig. \ref{C2H4-PES-one}. A smooth PEC can of course be reproduced by taking all the four frontier orbitals ($\sigma$, $\pi$, $\sigma^\ast$ and $\pi^\ast$) as active orbitals, i.e., CAS(4,4).
As pointed out by Malrieu long ago\cite{LMO-CASSCF}, the CAS(2,2) calculation can also produce a smooth PEC if localized $\sigma$ and $\sigma^\ast$ orbitals are used throughout.
This is indeed confirmed here by the present LMO-$\mathbbm{i}$CAS(2,2). Interestingly, the present CMO-$\mathbbm{i}$CAS(2,2), imposed by two pre-CMOs
formed by C$2\textrm{p}_x$ and H$1\textrm{s}$, can also pick up the two right active orbitals
and hence produces the same PEC as LMO-$\mathbbm{i}$CAS(2,2). The same findings hold also for the simultaneous breaking of two and even four \ce{C-H} bonds of ethylene, as
can be seen from Figs. \ref{C2H4-PES-two} to \ref{C2H4-PES-four}.
Although this example is very simple, it does reflect an important merit of $\mathbbm{i}$CAS:
the active space spanned by the prechosen AOs hold the same for all geometries.

\subsection{Electronic Structure of \ce{Co^{III}(diiminato)(NPh)}}\label{SecCo}
As a more stringent test of $\mathbbm{i}$CAS, we investigate the spin-state energetics of the low-coordinate imido complex \ce{Co^{III}(diiminato)(NPh)},
which is a simplified model of diamagnetic complex \ce{Co^{III}(nacnac)(NAd)} [NB: nacnac=an anion of 2,4-bis(2,6-dimethylphenylimido) pentane; Ad=1-adamantyl]\cite{CoN-2}.
The density functional theory study \cite{CoN-3} shows that there should be low-lying paramagnetic excited states besides the diamagnetic ground state, which has been calibrated by CAS(10,10)-based CASPT2\cite{CoN-1}.
The major concern here is how to construct an appropriate CAS. On the practical side,
the spin-free exact two-component (sf-X2C) Hamiltonian\cite{X2CSOC1,X2CSOC2} in conjunction with the ANO-RCC-VDZP basis sets (Co/5s4p2d1f; N,C/3s2p1d; H/2s1p)\cite{anorcc}
 (as used in Ref. \citenum{CoN-1}) is employed here. Twenty-nine core orbitals are frozen in the treatment of dynamic correlation, leaving in total 134 correlated
electrons. In the CASPT2 calculations (with the spin-free Douglas-Kroll-Hess (DKH) Hamiltonian\cite{DKH21989}), an imaginary level shift of 0.1 is used to remove intruder states.

The singlet ground state geometry  of \ce{Co^{III}(diiminato)(NPh)} was optimized with sf-X2C-PBE0/def2-SVP\cite{PBE0,DEF2-SVP},
 without symmetry restrictions. As can be seen from Fig. \ref{fig.Co-N}, the Co atom is located almost at the center of the molecular skeleton, the imido nitrogen (N$_{\textrm{imido}}$) is on the \textit{y} axis,
 the 1,3-propanediiminato and phenylimido groups are in the \textit{xy} plane, whereas the two terminal phenyl groups are perpendicular to the \textit{xy} plane (see Supporting Information for
 the Cartesian coordinates). The crucial coordinate of \ce{Co^{III}(diiminato)(NPh)} is the \ce{Co-N}$_{\textrm{imido}}$ bond, which was optimized to be 1.612 {\AA},
 very close to the CASPT2 value of 1.632 {\AA}\cite{CoN-1}.

\begin{figure}
\centering
\includegraphics[width=0.6\textwidth]{figures//Co-N}
\caption{Fragmentation of \ce{Co^{III}(diiminato)(NPh)}.}
\label{fig.Co-N}
\end{figure}

\begin{figure}
\centering
\includegraphics[width=1.0\textwidth]{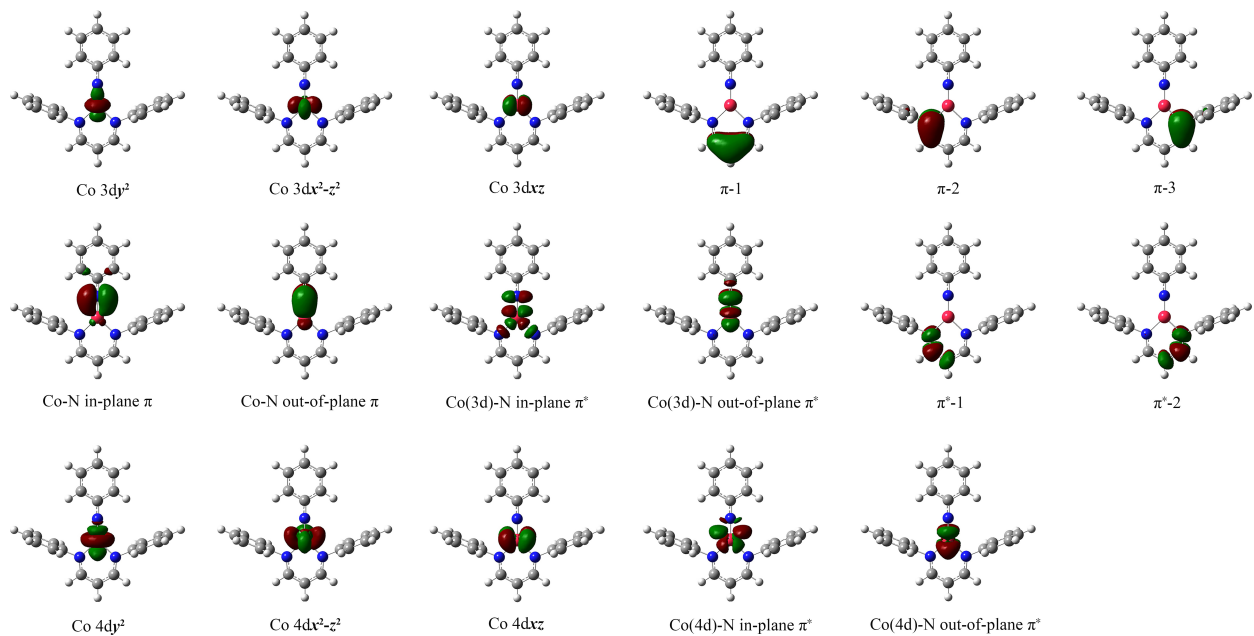}
\caption{Automatically selected initial 17 active LMOs of singlet \ce{Co^{III}(diiminato)(NPh)}. Co-N is the shorthand notation of \ce{Co-N$_{\textrm{imido}}$}.}
\label{LMO-CAS(16,17)}
\end{figure}

To initiate the ``top-down, least-change'' localization procedure (see Sec. \ref{preLMO}),  the molecule is partitioned into four fragments (see Fig. \ref{fig.Co-N}), with the dangling bonds saturated with the modified\cite{iOI} PHO (projected hybrid orbital) boundary atoms\cite{PHO}. While subsystems 1-3 are obviously closed-shell systems,
subsystem 4 can be in singlet, triplet or quintet. After the HF/ROHF calculations of
these subsystems, the orthonormal pFLMOs are generated\cite{F2M,ACR-FLMO}, which are taken as the initial guess for the HF/ROHF calculations of the whole systems.
The FLMOs are finally generated according to the ``top-down, least-change'' scheme. Since such FLMOs
have one-to-one correspondence with the pFLMOs, the active orbitals can be chosen simply from the subsystem pFLMOs. However, to make the $\mathbbm{i}$CAS calculations
``more difficult'', we still start with valence AOs. Looking at the chemical structure of the system, it should be easy to guess that, apart from
the five 3d orbitals of Co, the five 2p$_z$ orbitals of the C and N atoms of the 5-membered ring as well as the 2p$_x$ and 2p$_z$ of N$_{\textrm{imido}}$
are most relevant for the chemical environment of Co, since the 2p$_y$ orbital of N$_{\textrm{imido}}$ participates in the bonding with the 6-membered ring.
In addition to these valence AOs, the five 4d orbitals of Co can further be included to account for the so-called ``double d-shell effect''
resulting from physical intuitions\cite{double-shell}.
In essence, putting Co 4d into the active space amounts to shifting the inaccurate, perturbative treatment of their
radial correlation effect to the exact, variational treatment. We then have in total 17 valence AOs (and hence 17 pre-LMOs/CMOs) along with 16 active electrons
(6 for Co$^{3+}$ (3d$^6$), 6 for the anionic 5-membered ring, and 4 for $:$N$_{\textrm{imido}}^{2-}$)
according to minimal chemical and physical intuitions. Taking the 8 doubly occupied and 9 unoccupied pre-LMOs as probes, precisely
17 HF-LMOs (see Fig. \ref{LMO-CAS(16,17)}) can be selected as initial guess for $\mathbbm{i}$CAS(16,17).
As can be seen from Tables \ref{OverlapOccLMO-1th-iCAS(16,17)} and \ref{OverlapvirLMO-1th-iCAS(16,17)},
the large absolute overlaps between the pre-LMOs and HF-LMOs do support the selection. The same findings also hold for the triplet and quintet states (see
Supporting Information). Nevertheless, the CAS(16,17) calculations are too expensive (which, however, can readily be
performed with the iCISCF algorithm\cite{iCISCF} that employs the selected iterative configuration interaction (iCI)\cite{iCI,iCIPT2,iCIPT2New} as the CAS solver).
The first attempt to reduce CAS(16,17) is to exclude the five 2p$_z$ orbitals of the 5-membered ring, thereby leading to CAS(10,12).
As a matter of fact, this amounts just to deleting from Fig. \ref{LMO-CAS(16,17)} as well as Tables \ref{OverlapOccLMO-1th-iCAS(16,17)} and \ref{OverlapvirLMO-1th-iCAS(16,17)}
the $\pi-1$, $\pi-2$, $\pi-2$, $\pi^\ast-1$, and $\pi^\ast-2$ orbitals that are
localized on the 5-membered ring.
It was also attempted\cite{CoN-1} to further remove the  4d$_{xy}$ and 4d$_{yz}$ orbitals from CAS(10,12), for the 3d$_{y^2}$, 3d$_{x^2-z^2}$ and 3d$_{xz}$
orbitals are doubly occupied in the singlet ground state, such that only 4d$_{y^2}$, 4d$_{x^2-z^2}$ and 4d$_{xz}$ are relevant for the ``double d-shell effect''.
A more dramatic simplification is to exclude all the 4d orbitals from CAS(10,12), thereby leading to CAS(10,7).
These two cases amount to deleting from Fig. \ref{LMO-CAS(16,17)} the last two and five orbitals involving Co 4d, respectively.

Having prepared the initial guess, two-state averaged $\mathbbm{i}$CAS (SA2-$\mathbbm{i}$CAS) calculations are performed for each spin state.
As can be seen from Table \ref{OverlapIFLMO}, the converged $\mathbbm{i}$CAS(10,12) orbitals are indeed very close to the initial ones. Again, the same
findings also  hold for the triplet and quintet states (see Supporting Information). For completeness,
the leading configurations of the CAS wave functions are given in Table \ref{Table.Co-CSF}.
The vertical excitation energies calculated by $\mathbbm{i}$CAS, CASPT2, NEVPT2, and SDSPT2 are documented in Table \ref{Table.Co-SDSPT2}.
It can first be seen that, at the CASSCF level, CAS(10,7) is far from being sufficient: it even predicts that the two nearly degenerate $^3A$ states
are lower than the lowest singlet state by 0.3 eV; CAS(10,10) is also imperfect: although it does predict correctly the ground state\cite{CoN-2}, the excitation energies
of the remaining states are too high, by 0.3 to 0.6 eV. In contrast, CAS(10,12) is essentially saturated, with the predicted excitation energies differing
 from the CAS(16,17) ones by at most 0.15 eV. It can also be seen that the imperfections in CAS(10,10) and even CAS(10,7) are essentially removed by the
 subsequent CASPT2/NEVPT2/SDSPT2 treatment of dynamic correlation, indicating that the ``double d-shell effect'' plays a very minor role here. As a matter of fact,
 there exists very little dynamic correlation (ca. 0.1 eV) beyond CAS(10,12), as compared with CAS(10,12)-NEVPT2/SDSPT2.
 This can be explained by the fact that the spin states investigated here all arise from spin flips within the same Co 3d manifold (cf. Table \ref{Table.Co-CSF}), for which dynamic correlation
 effects are largely cancelled out for the relative energies. The situation may be different for states involving changes in the occupations of $n$d (and $n$f) orbitals,
 where substantial differential correlation effects may be experienced. Note also that there exist substantial differences (ca. 0.4 eV) between the present and the previous\cite{CoN-1}
 CAS(10,10)-CASPT2 excitation energies for the two $^5A$ states. Although the same CASPT2 code\cite{OpenMolcas} is used in both cases, the present
 calculation takes the SA2-iCAS(10,10) LMOs as input.

\begin{figure}
\begin{tabular}{ccc}
\resizebox{0.45\textwidth}{!}{\includegraphics{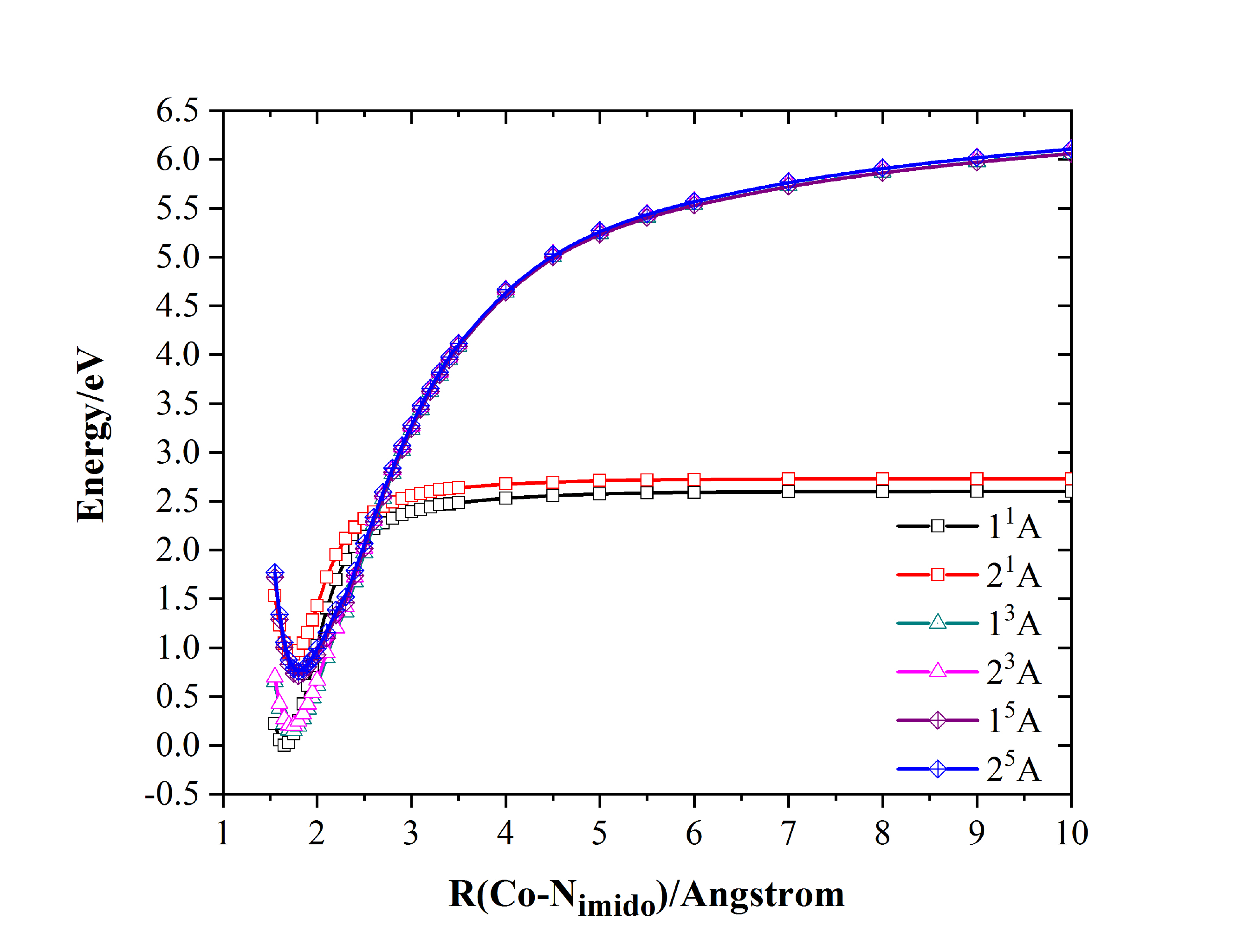}}&
\resizebox{0.45\textwidth}{!}{\includegraphics{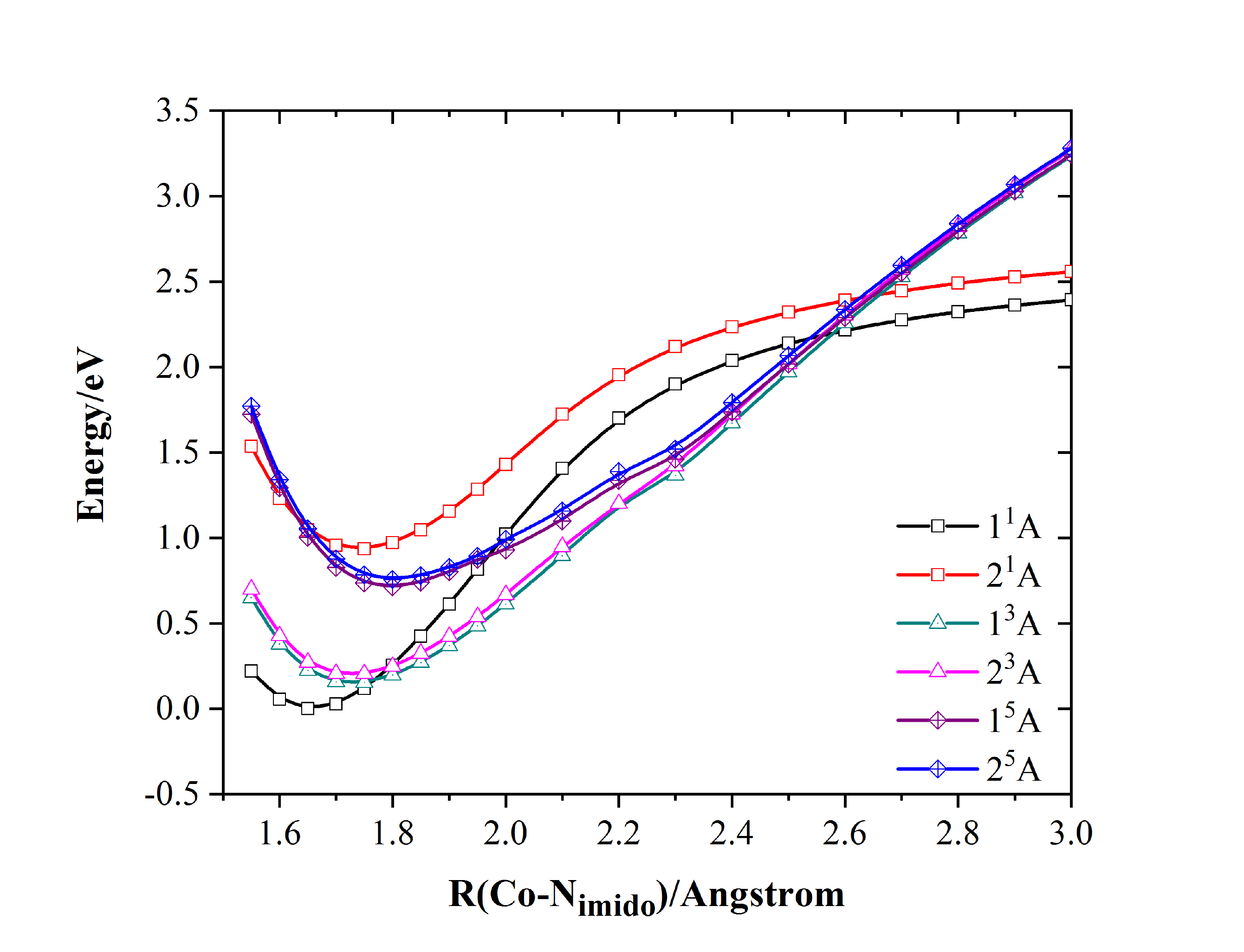}}&\\
(a) full   & (b)  zoom-in  \\
\end{tabular}
\caption{sf-X2C-SA2-$\mathbbm{i}$CAS(10,12)/ANO-RCC-VDZP potential energy curves for the rigid dissociation of Co-N$_{\textrm{imido}}$ in
\ce{Co^{III}(diiminato)(NPh)}. }
\label{fig.CoN-PECs}
\end{figure}

To further examine the efficacy of $\mathbbm{i}$CAS in maintaining the same CAS for all geometries, the rigid stretching of the Co-N$_{\textrm{imido}}$ bond (keeping the other atoms fixed) is considered.
The SA2-$\mathbbm{i}$CAS(10,12) PECs for the six states are plotted in Fig. \ref{fig.CoN-PECs}. The smoothness of the PECs is obvious.
While the Co-N$_{\textrm{imido}}$ bond in the $1^1A$ and $2^1A$ states starts to dissociate around 4.5 {\AA}, that in the four triplet and quintet states does not even up to 10 {\AA}.
It is also of interest to see that the four triplet and quintet states become degenerate beyond 2.3 {\AA}. Finally,
the equilibrium Co-N$_{\textrm{imido}}$ distances $R_e$ are given in Table \ref{Table.Co-N distance}.
Again, there exist significant differences (ca. 0.04 {\AA}) between the present and the previous\cite{CoN-1} CASPT2 values for $R_e$, due to the use of different CAS(10,10) orbitals.

\newpage

\begin{threeparttable}
  \caption{Absolute overlaps between the eight doubly occupied pre-LMOs and all doubly occupied molecular HF-LMOs for singlet \ce{Co^{III}(diiminato)(NPh)}}
  \tabcolsep=5pt
  \scriptsize
  \begin{tabular}[t]{@{}ccccccccccccc@{}}
   \hline\hline
   \diagbox[width=10em,height=3em,trim=l]{pre-LMO\tnote{a}}{active LMO} & 1  & 2  & 3  & 4  & 5 & 6 & 7 & 8 & 9 & $\cdots$\tnote{b} \\
    \hline
$b_\perp$	    &	0.960 	&	0.000 	&	0.000 	&	0.000 	&	0.000 	&	0.000 	&	0.000 	&	0.000 	&	0.000 	&	$\cdots$\\
$b_\parallel$	&	0.000 	&	0.954 	&	0.000 	&	0.000 	&	0.000 	&	0.000 	&	0.000 	&	0.000 	&	0.000 	&	$\cdots$\\
Co $\textrm{3d}_{xz}$   	&	0.000 	&	0.000 	&	0.999 	&	0.000 	&	0.000 	&	0.000 	&	0.000 	&	0.000 	&	0.000 	&	$\cdots$\\
Co $\textrm{3d}_{x^2-z^2}$	&	0.000 	&	0.000 	&	0.000 	&	0.721 	&	0.507 	&	0.000 	&	0.000 	&	0.000 	&	0.265 	&	$\cdots$\\
Co $\textrm{3d}_{y^2}$   	&	0.000 	&	0.000 	&	0.000 	&	0.550 	&	0.700 	&	0.000 	&	0.000 	&	0.000 	&	0.389 	&	$\cdots$\\
$\pi$-1     	&	0.032 	&	0.000 	&	0.000 	&	0.000 	&	0.000 	&	0.990 	&	0.000 	&	0.000 	&	0.000 	&	$\cdots$\\
$\pi$-2     	&	0.000 	&	0.000 	&	0.000 	&	0.000 	&	0.000 	&	0.000 	&	0.968 	&	0.000 	&	0.000 	&	$\cdots$\\
$\pi$-3     	&	0.000 	&	0.000 	&	0.000 	&	0.000 	&	0.000 	&	0.000 	&	0.000 	&	0.968 	&	0.000 	&	$\cdots$\\
	&	0.923\tnote{c} 	&	0.910\tnote{c} 	&	0.999\tnote{c} 	&	0.822\tnote{c} 	&	0.747\tnote{c} 	&	0.980\tnote{c} 	&	0.937\tnote{c} 	&	0.937\tnote{c} 	&	0.221\tnote{c} 	&	$\cdots$\\
   \hline \hline
  \end{tabular}\label{OverlapOccLMO-1th-iCAS(16,17)}
  \begin{tablenotes}
  \item [a] $b_\parallel$ and $b_\perp$ denote \ce{Co(3d)-N$_{\textrm{imido}}$} in plane $\pi$ and out-of-plane $\pi$; $\pi$-1, $\pi$-2 and $\pi$-3 correspond to localized $\pi$ of 1,3-propanediiminato.
  \item [b] Columns (core orbitals) with values smaller than 0.059.
  \item [c] Occupation number.
  \end{tablenotes}
\end{threeparttable}
\newpage
\begin{threeparttable}
  \caption{Absolute overlaps between the nine unoccupied pre-LMOs and all unoccupied molecular HF-LMOs for singlet \ce{Co^{III}(diiminato)(NPh)}}
  \tabcolsep=4pt
  \scriptsize
  \begin{tabular}[t]{@{}ccccccccccccc@{}}
   \hline\hline
   \diagbox[width=10em,height=3em,trim=l]{pre-LMO\tnote{a}}{active LMO} & 1 & 2 & 3 & 4 & 5 & 6 & 7 & 8 & 9 & 10 & $\cdots$\tnote{b} \\
    \hline
$b_\perp^\ast$	            &	0.861 	&	0.000 	&	0.000 	&	0.000 	&	0.000 	&	0.158 	&	0.000 	&	0.000 	&	0.000 	& 0.224	& $\cdots$\\
$b_\parallel^\ast$       	&	0.000 	&	0.742 	&	0.000 	&	0.000 	&	0.000 	&	0.000 	&	0.161 	&	0.000 	&	0.000 	& 0.000	& $\cdots$\\
Co $\textrm{4d}_{y^2}$	                &	0.000 	&	0.000 	&	0.599 	&	0.395 	&	0.000 	&	0.000 	&	0.000 	&	0.000 	&	0.000 	& 0.000	& $\cdots$\\
Co $\textrm{4d}_{x^2-z^2}$   	        &	0.000 	&	0.000 	&	0.452 	&	0.724 	&	0.000 	&	0.032 	&	0.000 	&	0.000 	&	0.000 	& 0.187	& $\cdots$\\
Co $\textrm{4d}_{xz}$               	&	0.000 	&	0.000 	&	0.000 	&	0.000 	&	0.913 	&	0.000 	&	0.000 	&	0.000 	&	0.000 	& 0.179	& $\cdots$\\
$\tilde{b}_\perp^\ast$	    &	0.130 	&	0.000 	&	0.000 	&	0.084 	&	0.000 	&	0.901 	&	0.000 	&	0.000 	&	0.000 	& 0.105	& $\cdots$\\
$\tilde{b}_\parallel^\ast$	&	0.000 	&	0.392 	&	0.000 	&	0.000 	&	0.000 	&	0.000 	&	0.778 	&	0.000 	&	0.000 	& 0.000	& $\cdots$\\
$\pi^\ast$-1	            &	0.000 	&	0.000 	&	0.000 	&	0.000 	&	0.000 	&	0.000 	&	0.000 	&	0.866 	&	0.032 	& 0.032	& $\cdots$\\
$\pi^\ast$-2             	&	0.000 	&	0.000 	&	0.000 	&	0.045 	&	0.000 	&	0.000 	&	0.000 	&	0.032 	&	0.866 	& 0.032	& $\cdots$\\
&	0.758\tnote{c} 	&	0.705\tnote{c} 	&	0.563\tnote{c} 	&	0.689\tnote{c} 	&	0.833\tnote{c} 	&	0.838\tnote{c} 	&	0.632\tnote{c} 	&	0.751\tnote{c} 	&	0.751\tnote{c} 	&	0.130\tnote{c} 	&	$\cdots$\\
   \hline \hline
  \end{tabular}\label{OverlapvirLMO-1th-iCAS(16,17)}
  \begin{tablenotes}
  \item [a] $b_\parallel^\ast$ and $b_\perp^\ast$ denote \ce{Co(3d)-N$_{\textrm{imido}}$} in plane and out-of-plane $\pi^\ast$; $\tilde{b}_\parallel^\ast$ and $\tilde{b}_\perp^\ast$ denote \ce{Co(4d)-N$_{\textrm{imido}}$} in plane and out-of-plane $\pi^\ast$; $\pi^\ast$-1 and $\pi^\ast$-2 correspond to localized $\pi^\ast$ of 1,3-propanediiminato.
  \item [b] Columns (virtual orbitals) with values smaller than 0.073.
  \item [c] Occupation number.
  \end{tablenotes}
\end{threeparttable}

\newpage
\begin{threeparttable}
  \caption{Absolute overlaps between the initial and final SA2-$\mathbbm{i}$CAS(10,12) LMOs for singlet \ce{Co^{III}(diiminato)(NPh)}}
  \tabcolsep=5pt
 \scriptsize
  \begin{tabular}[t]{@{}ccccccccccccc@{}}
    \hline \hline
    \diagbox[width=6em,height=3em,trim=l]{initial}{final}  & 1 & 2 & 3 & 4 & 5 & 6 & 7 & 8 & 9 & 10 & 11 & 12 \\
    \hline
Co $\textrm{3d}_{y^2}$          & \textbf{0.874}  & 0.000  & 0.001  & 0.000  & 0.000  & 0.000  & 0.000  & 0.000  & 0.014  & 0.000  & 0.000  & 0.000  \\
Co $\textrm{3d}_{x^2-z^2}$          & 0.001  & \textbf{0.911}  & 0.000  & 0.000  & 0.000  & 0.000  & 0.000  & 0.000  & 0.002  & 0.000  & 0.000  & 0.003  \\
Co $\textrm{3d}_{xz}$          & 0.001  & 0.000  & \textbf{0.998}  & 0.000  & 0.001  & 0.000  & 0.000  & 0.000  & 0.000  & 0.005  & 0.000  & 0.000  \\
$b_\parallel$          & 0.000  & 0.000  & 0.000  & \textbf{0.978}  & 0.000  & 0.000  & 0.002  & 0.000  & 0.000  & 0.000  & 0.008  & 0.000  \\
$b_\perp$          & 0.000  & 0.000  & 0.000  & 0.000  & \textbf{0.989}  & 0.008  & 0.000  & 0.012  & 0.000  & 0.000  & 0.000  & 0.005  \\
$b_\perp^\ast$          & 0.000  & 0.000  & 0.000  & 0.008  & 0.000  & \textbf{0.886}  & 0.000  & 0.014  & 0.000  & 0.000  & 0.000  & 0.021  \\
$b_\parallel^\ast$          & 0.000  & 0.002  & 0.000  & 0.000  & 0.000  & 0.000  & \textbf{0.894}  & 0.001  & 0.000  & 0.000  & 0.121  & 0.001  \\
Co $\textrm{4d}_{y^2}$          & 0.000  & 0.000  & 0.014  & 0.000  & 0.002  & 0.000  & 0.000  & \textbf{0.715}  & 0.003  & 0.001  & 0.000  & 0.017  \\
Co $\textrm{4d}_{x^2-z^2}$          & 0.000  & 0.000  & 0.000  & 0.005  & 0.003  & 0.021  & 0.001  & 0.040  & \textbf{0.863}  & 0.001  & 0.000  & 0.017  \\
Co $\textrm{4d}_{xz}$          & 0.005  & 0.000  & 0.000  & 0.000  & 0.000  & 0.000  & 0.000  & 0.000  & 0.001  & \textbf{0.892}  & 0.000  & 0.001  \\
$\tilde{b}_\perp^\ast$          & 0.000  & 0.000  & 0.000  & 0.012  & 0.000  & 0.014  & 0.001  & 0.003  & 0.003  & 0.000  & \textbf{0.876}  & 0.040  \\
$\tilde{b}_\parallel^\ast$          & 0.000  & 0.008  & 0.000  & 0.000  & 0.000  & 0.000  & 0.121  & 0.003  & 0.000  & 0.000  & 0.000  & \textbf{0.753}  \\
    \hline \hline
  \end{tabular}\label{OverlapIFLMO}
\end{threeparttable}

\newpage

\begin{threeparttable}
  \caption{Leading configurations of HF/ROHF and $\mathbbm{i}$CAS calculations of \ce{Co^{III}(diiminato)(NPh)}}
  \tabcolsep=5pt
 \scriptsize
  \begin{tabular}[t]{@{}clcccccccccccccc@{}}
    \hline \hline
    state & \multicolumn{1}{c}{main configuration\tnote{a}} & energy (eV)/weight\tnote{b} \\
    \hline
                                           \multicolumn{3}{c}{HF/ROHF}\\
    1$^1$A & $(\textrm{3d}_{y^2})^2(\textrm{3d}_{x^2-z^2})^2(\textrm{3d}_{xz})^2(b_\parallel)^2(b_\perp)^2$   & 0.00 \\
    1$^3$A & $(\textrm{3d}_{y^2})^2(\textrm{3d}_{x^2-z^2})^2(\textrm{3d}_{xz})^2(b_\parallel)^2
    (b_\perp)^1(b_\perp^\ast)^1$ & -1.40 \\
    1$^5$A & $(\textrm{3d}_{y^2})^1(\textrm{3d}_{x^2-z^2})^2(\textrm{3d}_{xz})^1(b_\parallel)^2(b_\perp)^2
    (b_\perp^\ast)^1(b_\parallel^\ast)^1$ & -3.34 \\
                                           \multicolumn{3}{c}{$\mathbbm{i}$CAS(10,7)}\\
    1$^1$A & $(\textrm{3d}_{y^2})^2(\textrm{3d}_{x^2-z^2})^2(\textrm{3d}_{xz})^2(b_\parallel)^2(b_\perp)^2
    (b_\perp^\ast)^0(b_\parallel^\ast)^0$  & 0.761 \\
    2$^1$A & $(\textrm{3d}_{y^2})^2(\textrm{3d}_{x^2-z^2})^2(\textrm{3d}_{xz})^2(b_\parallel)^2(b_\perp)^1
    (b_\perp^\ast)^1(b_\parallel^\ast)^0$  & 0.804 \\
    1$^3$A & $(\textrm{3d}_{y^2})^2(\textrm{3d}_{x^2-z^2})^1(\textrm{3d}_{xz})^2(b_\parallel)^2(b_\perp)^2
    (b_\perp^\ast)^1(b_\parallel^\ast)^0$  & 0.738 \\
    2$^3$A & $(\textrm{3d}_{y^2})^2(\textrm{3d}_{x^2-z^2})^2(\textrm{3d}_{xz})^1(b_\parallel)^2(b_\perp)^2
    (b_\perp^\ast)^1(b_\parallel^\ast)^0$  & 0.740 \\
    1$^5$A & $(\textrm{3d}_{y^2})^1(\textrm{3d}_{x^2-z^2})^1(\textrm{3d}_{xz})^2(b_\parallel)^2(b_\perp)^2
    (b_\perp^\ast)^1(b_\parallel^\ast)^1$   & 0.820 \\
    2$^5$A & $(\textrm{3d}_{y^2})^1(\textrm{3d}_{x^2-z^2})^2(\textrm{3d}_{xz})^1(b_\parallel)^2(b_\perp)^2
    (b_\perp^\ast)^1(b_\parallel^\ast)^1$   & 0.822 \\
                                           \multicolumn{3}{c}{$\mathbbm{i}$CAS(10,10)}\\
    1$^1$A & $(\textrm{3d}_{y^2})^2(\textrm{3d}_{x^2-z^2})^2(\textrm{3d}_{xz})^2(b_\parallel)^2(b_\perp)^2
    (b_\perp^\ast)^0(b_\parallel^\ast)^0(\textrm{4d}_{y^2})^0(\textrm{4d}_{x^2-z^2})^0(\textrm{4d}_{xz})^0$  & 0.771 \\
    2$^1$A & $(\textrm{3d}_{y^2})^2(\textrm{3d}_{x^2-z^2})^2(\textrm{3d}_{xz})^2(b_\parallel)^2(b_\perp)^1
    (b_\perp^\ast)^1(b_\parallel^\ast)^0(\textrm{4d}_{y^2})^0(\textrm{4d}_{x^2-z^2})^0(\textrm{4d}_{xz})^0$  & 0.812 \\
    1$^3$A & $(\textrm{3d}_{y^2})^2(\textrm{3d}_{x^2-z^2})^1(\textrm{3d}_{xz})^2(b_\parallel)^2(b_\perp)^2
    (b_\perp^\ast)^1(b_\parallel^\ast)^0(\textrm{4d}_{y^2})^0(\textrm{4d}_{x^2-z^2})^0(\textrm{4d}_{xz})^0$  & 0.768 \\
    2$^3$A & $(\textrm{3d}_{y^2})^2(\textrm{3d}_{x^2-z^2})^2(\textrm{3d}_{xz})^1(b_\parallel)^2(b_\perp)^2
    (b_\perp^\ast)^1(b_\parallel^\ast)^0(\textrm{4d}_{y^2})^0(\textrm{4d}_{x^2-z^2})^0(\textrm{4d}_{xz})^0$  & 0.769 \\
    1$^5$A & $(\textrm{3d}_{y^2})^1(\textrm{3d}_{x^2-z^2})^1(\textrm{3d}_{xz})^2(b_\parallel)^2(b_\perp)^2
    (b_\perp^\ast)^1(b_\parallel^\ast)^1(\tilde{b}_\perp^\ast)^0(\textrm{4d}_{x^2-z^2})^0(\textrm{4d}_{xz})^0$   & 0.909 \\
    2$^5$A & $(\textrm{3d}_{y^2})^1(\textrm{3d}_{x^2-z^2})^2(\textrm{3d}_{xz})^1(b_\parallel)^2(b_\perp)^2
    (b_\perp^\ast)^1(b_\parallel^\ast)^1(\tilde{b}_\perp^\ast)^0(\textrm{4d}_{x^2-z^2})^0(\textrm{4d}_{xz})^0$   & 0.904 \\
                                           \multicolumn{3}{c}{$\mathbbm{i}$CAS(10,12)}\\
    1$^1$A & $(\textrm{3d}_{y^2})^2(\textrm{3d}_{x^2-z^2})^2(\textrm{3d}_{xz})^2(b_\parallel)^2(b_\perp)^2
    (b_\perp^\ast)^0(b_\parallel^\ast)^0(\textrm{4d}_{y^2})^0(\textrm{4d}_{x^2-z^2})^0(\textrm{4d}_{xz})^0
    (\tilde{b}_\perp^\ast)^0(\tilde{b}_\parallel^\ast)^0$  & 0.742 \\
    2$^1$A & $(\textrm{3d}_{y^2})^2(\textrm{3d}_{x^2-z^2})^2(\textrm{3d}_{xz})^2(b_\parallel)^2(b_\perp)^1
    (b_\perp^\ast)^1(b_\parallel^\ast)^0(\textrm{4d}_{y^2})^0(\textrm{4d}_{x^2-z^2})^0(\textrm{4d}_{xz})^0
    (\tilde{b}_\perp^\ast)^0(\tilde{b}_\parallel^\ast)^0$  & 0.778 \\
    1$^3$A & $(\textrm{3d}_{y^2})^2(\textrm{3d}_{x^2-z^2})^1(\textrm{3d}_{xz})^2(b_\parallel)^2(b_\perp)^2
    (b_\perp^\ast)^1(b_\parallel^\ast)^0(\textrm{4d}_{y^2})^0(\textrm{4d}_{x^2-z^2})^0(\textrm{4d}_{xz})^0
    (\tilde{b}_\perp^\ast)^0(\tilde{b}_\parallel^\ast)^0$  & 0.771 \\
    2$^3$A & $(\textrm{3d}_{y^2})^2(\textrm{3d}_{x^2-z^2})^2(\textrm{3d}_{xz})^1(b_\parallel)^2(b_\perp)^2
    (b_\perp^\ast)^1(b_\parallel^\ast)^0(\textrm{4d}_{y^2})^0(\textrm{4d}_{x^2-z^2})^0(\textrm{4d}_{xz})^0
    (\tilde{b}_\perp^\ast)^0(\tilde{b}_\parallel^\ast)^0$  & 0.760 \\
    1$^5$A & $(\textrm{3d}_{y^2})^1(\textrm{3d}_{x^2-z^2})^1(\textrm{3d}_{xz})^2(b_\parallel)^2(b_\perp)^2
    (b_\perp^\ast)^1(b_\parallel^\ast)^1(\textrm{4d}_{y^2})^0(\textrm{4d}_{x^2-z^2})^0(\textrm{4d}_{xz})^0
    (\tilde{b}_\perp^\ast)^0(\tilde{b}_\parallel^\ast)^0$   & 0.834 \\
    2$^5$A & $(\textrm{3d}_{y^2})^1(\textrm{3d}_{x^2-z^2})^2(\textrm{3d}_{xz})^1(b_\parallel)^2(b_\perp)^2
    (b_\perp^\ast)^1(b_\parallel^\ast)^1(\textrm{4d}_{y^2})^0(\textrm{4d}_{x^2-z^2})^0(\textrm{4d}_{xz})^0
    (\tilde{b}_\perp^\ast)^0(\tilde{b}_\parallel^\ast)^0$   & 0.832 \\
    \hline \hline
  \end{tabular}\label{Table.Co-CSF}
  \begin{tablenotes}
  \item [a] $b_\parallel$ ($b_\parallel^\ast$) and $b_\perp$ ($b_\perp^\ast$) denote \ce{Co(4d)-N$_{\textrm{imido}}$} in plane and out-of-plane $\pi$ ($\pi^\ast$), respectively;
   $\tilde{b}_\perp^\ast$ denotes \ce{Co(4d)-N$_{\textrm{imido}}$} out-of-plane $\pi^\ast$ with $\textrm{4d}_{yz}$ as the major.
   \item [b] HF/ROHF total energies relative to the singlet (-2361.437053 a.u.); weights of leading configurations for CASSCF.
  \end{tablenotes}
\end{threeparttable}

\newpage
\begin{threeparttable}
  \caption{Vertical excitation energies (in eV) of \ce{Co^{III}(diiminato)(NPh)}
  calculated at the singlet ground state geometry (see Supporting Information)}
  \tabcolsep=3pt
 \scriptsize
  \begin{tabular}[t]{@{}cccccccccccccccccccccccc@{}}
    \hline \hline
    &\multicolumn{4}{c}{$\mathbbm{i}$CAS\tnote{a}} &   &\multicolumn{3}{c}{CASPT2}&  &\multicolumn{3}{c}{NEVPT2} &  &\multicolumn{3}{c}{SDSPT2}\\
    \cline{2-5}\cline{7-9}\cline{11-13}\cline{15-17}
    state & CAS1 & CAS2 & CAS3 & CAS4\tnote{b} &   & CAS1 & CAS2\tnote{c} & CAS3 &    & CAS1  & CAS2 & CAS3   &   & CAS1 & CAS2 & CAS3 \\
    \hline
    1$^1$A   & 0.00  & 0.00  & 0.00  & 0.00 &   & 0.00  & 0.00(0.00)  & 0.00  &    & 0.00  & 0.00  & 0.00  &   & 0.00  & 0.00  & 0.00  \\
    2$^1$A   & 0.93  & 1.77  & 1.15  & 1.19 &   & 1.01  & 0.88(0.83)  & 1.01  &    & 1.14  & 0.86  & 1.03  &   & 1.14  & 0.89  & 1.04  \\
    1$^3$A   &-0.37  & 0.57  & 0.30  & 0.43 &   & 0.34  & 0.24(0.11)  & 0.27  &    & 0.55  & 0.33  & 0.33  &   & 0.52  & 0.34  & 0.34  \\
    2$^3$A   &-0.31  & 0.63  & 0.35  & 0.50 &   & 0.38  & 0.29(0.15)  & 0.32  &    & 0.60  & 0.37  & 0.38  &   & 0.57  & 0.37  & 0.38  \\
    1$^5$A   & 0.26  & 1.65  & 1.19  & 1.35 &   & 1.01  & 0.91(0.59)  & 1.03  &    & 1.31  & 1.13  & 1.31  &   & 1.28  & 1.15  & 1.31  \\
    2$^5$A   & 0.32  & 1.72  & 1.24  & l.36 &   & 0.95  & 0.95(0.59)  & 1.08  &    & 1.37  & 1.14  & 1.35  &   & 1.34  & 1.16  & 1.35  \\
    \hline \hline
  \end{tabular}\label{Table.Co-SDSPT2}
  \begin{tablenotes}
  \item [a] CAS1, CAS2, CAS3 and CAS4 denote CAS(10,7), CAS(10,10), CAS(10,12) and CAS(16,17), respectively.
  \item [b] Calculated with the iCISCF approach\cite{iCISCF}.
  \item [c] Values in parentheses are results from Ref. \citenum{CoN-1}.
  \end{tablenotes}
\end{threeparttable}


\newpage

\newpage
\begin{threeparttable}
  \caption{Equilibrium distances (in {\AA}) for the rigid dissociation of Co-N$_{\textrm{imido}}$ in 
\ce{Co^{III}(diiminato)(NPh)}}
  \tabcolsep=30pt
 \scriptsize
  \begin{tabular}[t]{@{}cccccccccccccccc@{}}
    \hline \hline
    state & $\mathbbm{i}$CAS(10,12)  & CASPT2 & NEVPT2 & SDSPT2 \\
    \hline
    1$^1$A & 1.660  & 1.605  & 1.602  & 1.602  \\
    2$^1$A & 1.752  & 1.697  & 1.697  & 1.697  \\
    1$^3$A & 1.743  & 1.651  & 1.649  & 1.651  \\
    2$^3$A & 1.743  & 1.651  & 1.649  & 1.651  \\
    1$^5$A & 1.798  & 1.706  & 1.743  & 1.743  \\
    2$^5$A & 1.798  & 1.706  & 1.743  & 1.743  \\
    \hline \hline
  \end{tabular}\label{Table.Co-N distance}
\end{threeparttable}

\section{Conclusions and outlook}\label{Conclusion}
An $\mathbbm{i}$CAS method has been proposed to serve as an initial step for describing strongly correlated systems of electrons.
It has several features: (1) the active orbitals can be selected automatically by starting with prechosen valence/core atomic/fragmental orbitals.
The only requirement lies in that such atomic/fragmental orbitals span a space that is sufficient for the target many-electron states,
which hold for most organic molecules or transition metal complexes. (2) The CASSCF orbitals can readily be localized in each iteration so as to
facilitate the matching of core, active and virtual subspaces. 
(3) The CAS can be maintained the same for all geometries, which is of vital importance for the scanning of potential energy surfaces.
(4) Further combined with the iterative vector interaction (iVI) approach\cite{iVI,iVI-TDDFT} for directly accessing interior roots of the CI eigenvalue problem,
the CASSCF calculation can be dictated to converge to a predesignated chare/spin configuration composed of some special motifs (e.g., metal
valence d/f orbitals, atomic core orbitals or aromatic rings), a point that is to be investigated in future though.

\section*{Acknowledgement}
This work was supported by 
the National Natural Science Foundation of China (Grant Nos. 21833001 and 21973054),
Natural Science Basic Research Plan of Shaanxi Province (Grant No. 2019JM-196),
Mountain Tai Climb Program of Shandong Province, and Key-Area Research and Development Program of Guangdong Province (Grant No. 2020B0101350001).

\section*{Data Availability Statement}
The Cartesian coordinates of the \ce{Co^{III}(diiminato)(NPh)} as well as various plots of the CASSCF orbitals
 are documented in the Supporting Information.

\clearpage
\newpage

\bibliography{FLMO-CASSCF}

\end{document}